\documentclass[twocolumn]{aastex631}

\usepackage{amsmath}
\usepackage[normalem]{ulem}

\newcommand{\arepo}{\textsc{arepo}}
\newcommand{\msol}{$\mathrm{M}_\odot$}
\newcommand{\kms}{km s$^{-1}$}

\newcommand{\rev}[2]{#2}

\begin{document}

\title{On the Origin of High-velocity Clouds in the Galaxy}

\author[0000-0001-9982-0241]{Scott Lucchini}
\affiliation{Center for Astrophysics $|$ Harvard \& Smithsonian, 60 Garden Street, Cambridge, MA 02138, USA}

\correspondingauthor{Scott Lucchini}
\email{scott.lucchini@cfa.harvard.edu}

\author[0000-0002-6800-5778]{Jiwon Jesse Han}
\affiliation{Center for Astrophysics $|$ Harvard \& Smithsonian, 60 Garden Street, Cambridge, MA 02138, USA}

\author[0000-0001-6950-1629]{Lars Hernquist}
\affiliation{Center for Astrophysics $|$ Harvard \& Smithsonian, 60 Garden Street, Cambridge, MA 02138, USA}

\author[0000-0002-1590-8551]{Charlie Conroy}
\affiliation{Center for Astrophysics $|$ Harvard \& Smithsonian, 60 Garden Street, Cambridge, MA 02138, USA}
 
\begin{abstract}

The origin of our Galaxy's high-velocity clouds (HVCs) remains a mystery after many decades of effort. In this paper, we use the TNG50 simulation of the IllustrisTNG project to identify cool, dense clouds that match observations of Galactic \rev{}{\ion{H}{1}} HVCs. We track these clouds back in time to determine their origin. \rev{}{For a TNG50 Milky Way-like galaxy, }we find that only \rev{}{17}\% of HVCs \rev{(by mass)}{} can be tracked directly to the disk, and \rev{36}{21}\% to \rev{satellite  accretion}{material stripped out of satellites}. The majority of HVCs (\rev{52}{62}\%) arise from warm and hot circumgalactic gas that cools through thermal instability. They then obtain their anomalous velocities through interactions with the turbulent circumgalactic medium. At TNG50 resolution, we do not see evidence for HVCs forming out of very low metallicity intergalactic material. Instead, low metallicity HVCs are most likely associated with satellites\rev{ accretion}{}. These results suggest that Galactic HVCs are highly heterogeneous in their origin, and can provide insight into the physical processes that shape the circumgalactic medium such as disk outflows, satellite accretion, and thermal instabilities.

\end{abstract}


\section{Introduction} \label{sec:intro}

The circumgalactic medium (CGM) plays a critical role in galaxy evolution -- it is the interface between a galaxy and its larger cosmological environment. The CGM also contains a substantial reservoir of gas that can continue to feed the galaxy and maintain star formation over Gyr timescales \citep{spitzer56,binney77,chevalier79,tumlinson17}.
This star formation and gas accretion subsequently induces strong feedback from supernovae and active galactic nuclei returning gas and energy to the CGM and beyond.
However, this process of accretion and outflow -- the ``baryon cycle'' --  remains poorly understood in detail. One of the most significant puzzles is how we can connect our theoretical models and simulations back to observations of the CGM \citep{peeples19,faucher-giguere23}.

Some of the earliest observations of circumgalactic gas come from our own Galaxy in the form of ``high-velocity clouds'' (HVCs; \citealt{vanwoerden57,muller67,wakker97,putman12}). As observations have improved we have uncovered more information about these elusive clouds \rev{}{that trace the inner CGM and disk-halo interface}, but our understanding of their properties is still limited. Modern \ion{H}{1} 21-cm emission surveys give us an excellent view of their warm, neutral gas structure and morphology \citep{kalberla06,moss13,westmeier18}, and absorption spectroscopy can give us information about their ionization state, composition, and distance \citep[e.g.,][]{lehner11,smoker11,fox14,fox16,richter17,lehner22}.  However, these sightlines are not as numerous as necessary to obtain strong constraints on the global properties of HVCs. \rev{}{Additionally, H$\alpha$ emission can provide further constraints on distances based on models of the MW ionizing radiation field \citep{bland-hawthorn99,putman03}.}

A fundamental mystery surrounding the HVC population of the Milky Way (MW) is its origins. Many models have been proposed, including \rev{satellite accretion}{gas from accreting satellites} \citep{olano08,richter18}, galactic fountain recycling \citep{habe80,bregman80,fox23},
accretion from the intergalactic medium \citep{keres09}, or the precipitation of small cloudlets \citep{field65,maller04,voit21,tripp22}. While any single model is unlikely to explain all of the HVCs in the Galaxy, it is important to understand the relative importance of each mechanism at play in the Galactic CGM. Cosmological hydrodynamical simulations pose an opportunity to tackle precisely this problem in a realistic galactic environment.

\begin{figure}
    \centering
    \includegraphics[width=\columnwidth]{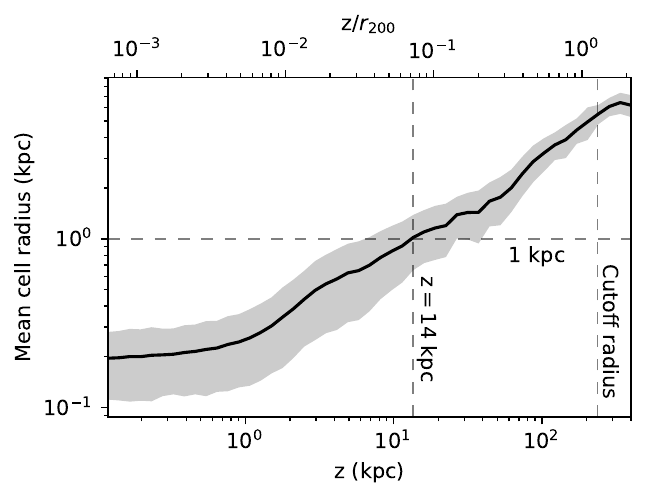}
    \caption{Mean gas cell spatial resolution as a function of $z$ height above the galactic disk. Distances and cell sizes are given in physical kpc.}
    \label{fig:resolution}
\end{figure}

The CGM is notoriously difficult to simulate due to its low density, extremely large volume, and high resolution requirements \citep{nelson16,vandevoort19,hummels19,faucher-giguere23}. It is important to capture both the cosmological context (e.g., merger histories, intergalactic medium enrichment) and the small-scale processes (e.g., turbulence, instabilities) to describe this complex environment \citep{lochhaas23}. For this reason, the IllustrisTNG cosmological box simulations provide an excellent avenue for exploration of the CGM. The TNG50 simulation \citep{nelson19,pillepich19} has high baryonic resolution and the publicly available subboxes provide high time cadence outputs perfect for the analysis of HVC origins.

Previous investigations with cosmological simulations have explored the origins of the CGM as a whole \citep{hafen19}, the cold component \citep{decataldo24}, or the high-velocity cold component \citep{ramesh23}. In this work, we offer a complementary view by identifying HVC analogs in the TNG50 simulation based on criteria specifically motivated by Galactic \ion{H}{1} observations. By tracking these HVCs into the past, we identify their origins, and investigate their relative contributions to the overall HVC population.

This paper is organized as follows. In Section~\ref{sec:methods} we describe the simulations and our analysis techniques for identifying and tracking the clouds. Section~\ref{sec:results} compiles our distribution of origins and explores the specific histories of several clouds. We discuss our results and conclude in Sections~\ref{sec:discussion} and \ref{sec:conclusions}.

\section{Methods} \label{sec:methods}


\subsection{The TNG Simulations} \label{sec:tng}

IllustrisTNG is a suite of high resolution cosmological box simulations using the magnetohydrodynamic code \arepo \, \citep{weinberger17,pillepich18}. The TNG suite consists of three volumes -- TNG50, TNG100, and TNG300 corresponding to simulation cubes of 50, 100, and 300 physical Mpc on a side \citep{nelson19b}. In this work, we use the TNG50-1 simulation which has mass resolutions for dark matter and gas of $4.5\times10^5$ and $8.5\times10^4$ \msol, respectively. TNG also produced several ``subboxes'' in which a smaller region of the simulation is output at much higher time cadence ($\sim$6 Myr). \texttt{subbox0} contains a region of space with moderate density which contains $\sim$ 6 MW-like galaxies. We select one such MW-like galaxy corresponding to subhalo ID 537941 at $z=0$ in the full TNG50-1 simulation. This is the only galaxy in the TNG50 MW/M31 sample that is included in a subbox from early times \citep{pillepich23}.

This galaxy has a total mass of $7.64\times10^{11}$ \msol\ at $z=0.5$, increasing to $1.02\times10^{12}$ \msol\ by $z=0$. Its stellar mass is $\sim3\times10^{10}$ \msol\ and its star formation rate is slightly higher than that of our own Galaxy at $\sim5$ \msol/yr \citep{bland-hawthorn16}.

To visualize the physical gas cell resolution, Figure~\ref{fig:resolution} shows the mean effective cell radius (the radius of a sphere with equivalent volume) as a function of height above the plane for our $z=0.25$ snapshot. The bottom axis shows height in physical kpc while the top axis shows the ratio of height over virial radius of the halo ($r_{200}=186$ kpc). We include all gas within 200 comoving kpc of the galaxy in our cold cloud analysis (the ``cutoff radius''; 237 physical kpc at $z=0.25$). While this resolution is not sufficient to resolve the smallest clouds, it does capture thermal instabilities and cloud growth on larger scales (see Section~\ref{sec:discussion}).

\begin{figure*}
    \centering
    \includegraphics[width=0.85\textwidth]{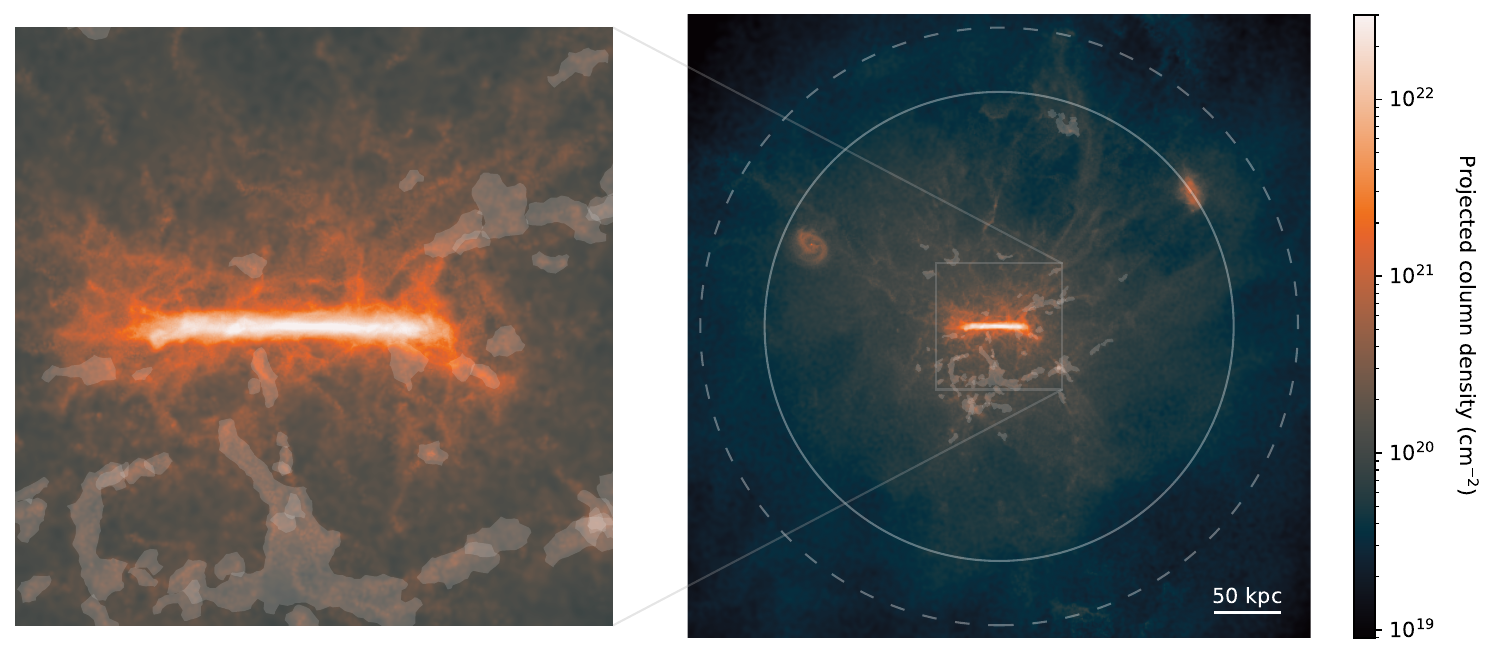}
    \caption{The locations of the identified HVC analogs overlaid on a projection of the galaxy at our final time ($z\sim0.25$). The solid circle in the right panel shows the virial radius of the galaxy ($r_{200}=186$~kpc) and the dashed circle shows the limit of our cloud identification region (``cutoff radius'', $r_\mathrm{cut}=200\;\mathrm{ckpc}=237\;\mathrm{pkpc}$).}
    \label{fig:cartesian}
\end{figure*}

\subsection{Cloud Identification} \label{sec:cloud_identification}

To identify contiguous groupings of cold gas cells, we regenerate the Voronoi mesh for the cells in a given snapshot. The \arepo\ output files contain gas particles that represent the positions of each Voronoi cell. Thus the Voronoi tessellation can be reconstructed from the snapshot output of the code. We then check each cell within 200 comoving kpc to see if it meets our ``cold cloud'' criteria. For this work, we are including any cells in which $T<10^{4.5}$ K. Using a recursive neighbor-finding algorithm, we can obtain a list of cell IDs which are contiguous and meet our temperature criteria.

This technique results in a list of ``clouds'' for each snapshot. Each cloud consists of one or more gas cells with $T<10^{4.5}$ K. We can then calculate mean properties for each of these clouds by averaging the properties of its constituent cells (e.g. position, velocity, mass, density, temperature, etc.). The detected cloud with the largest number of members will be the galactic disk and we exclude it from our analysis. This method is based on techniques published in \citet{nelson20} and \citet{ramesh23}.
A Cartesian projection of the galaxy with the identified clouds is shown in Figure~\ref{fig:cartesian}

\subsection{Cloud Tracking} \label{sec:cloud_tracking}

Once the clouds in each snapshot have been identified, we track them in time to generate a merger tree for each cloud (inspired by \citealt{jeffreson21}). For every cloud we propagate each of its member cells forward in time based on their current positions and velocities. We then compute the convex hull of the future cell positions and compare against all of the clouds in the next snapshot. If more than 10\% of the gas cells in any of the detected clouds in this next snapshot are contained within the convex hull of the propagated cloud, we link these two clouds.
The merger tree itself is generated using a graph matrix (specifically, the python implementation \texttt{NetworkX}\footnote{\url{https://networkx.org}}). Each cloud is a node in the graph and clouds are linked with graph ``edges.''

We then prune each connected component of the graph to determine a single evolutionary track for each cloud. By removing dead end branches and taking the largest clouds upon splitting, we are left with one cloud per snapshot that we can follow in the analysis below.

This is distinct from previous works in that we are tracking the observable cold cloud itself in time as opposed to using tracer particles to follow the motion of individual gas parcels. Due to gas mixing driven by radiative cooling \citep[e.g.,][]{gronke18}, the actual gas parcels within the cloud will change even if the global properties of the cloud are relatively constant. In this paper, we want to track these physical clouds to identify how their initial seeds formed. However each cloud is formed at a certain time and before that time, the above tracking method doesn't work.

The first snapshot in the evolutionary track of each cloud is its initial snapshot. Prior to this time, the cloud either has too few members or its temperature has risen enough that it is no longer identified using the algorithm described above (Section~\ref{sec:cloud_identification}). In order to track these clouds back in time beyond this point, we resort to two imperfect methods. The most natural choice would be to use the Monte-Carlo tracer particles that reside in the cloud at this initial snapshot \citep{genel13}. However, due to the overdiffusive nature of the tracers in \arepo, we have also tracked these cloud precursors using gas cell IDs. Obviously using the IDs does not consider the gas flux between cells during their evolution, however out in the CGM, we expect these fluxes to be relatively small and this will allow us to explore the nondiffusive limit when compared with the overdiffusive tracers. As we show below, we obtain consistent results when using both methods.

\begin{figure*}
    \centering
    \includegraphics[width=0.9\textwidth]{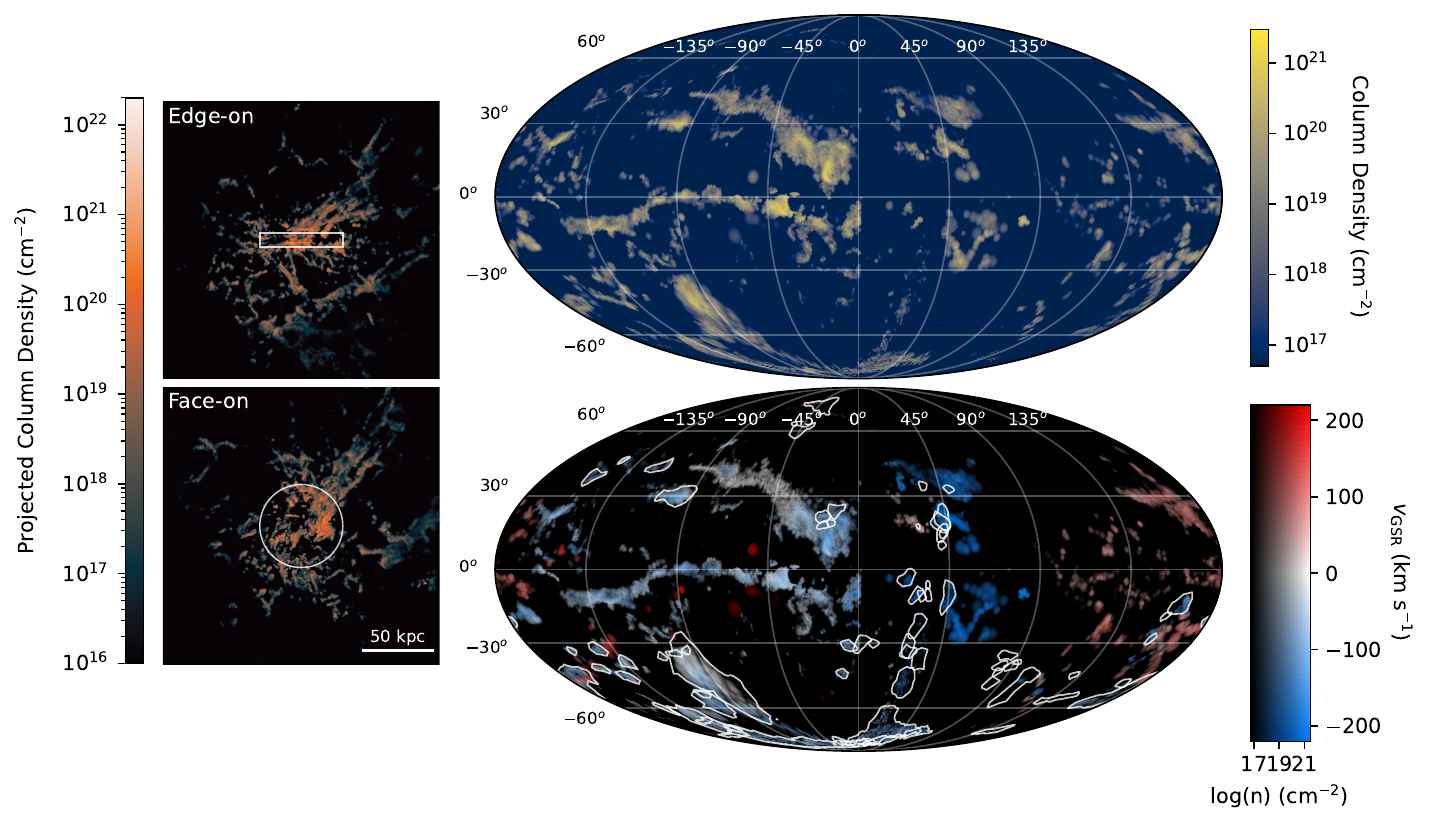}
    \caption{The distribution of high velocity gas. On the left, we show two perspectives in cartesian coordinates with the approximate extent of the disk outlined in white. On the right, we show the high velocity gas projected onto the sky assuming a solar-like position within the galaxy. The several large complexes near the plane are physically connected to the disk and are not included in our HVC analysis. The top panel shows column density and the bottom shows GSR velcity with the identified HVCs outlined in white. \rev{}{The several large complexes near the equator are not identified as HVCs because they are anomalous velocity gas connected to the galactic disk (due to the disk's warp). There are several other small clouds that are not identified because they are too small.}}
    \label{fig:onsky}
\end{figure*}

\subsection{HVC Analogs} \label{sec:hvc_analogs}

In order to determine those clouds that are most similar to HVCs observed around the MW, we use a mock observation technique following \citet{westmeier18}. We first determine the expected line of sight velocity due to disk rotation using the following formula.
\begin{equation}
    v_\mathrm{disk}(\ell,b,R) = \left(v_\mathrm{rot}(R)\frac{R_0}{R} - v_\mathrm{rot}(R_0)\right)\sin(\ell)\cos(b)
\end{equation}
Here, $\ell$ and $b$ are Galactic longitude and latitude, respectively, $R$ is cylindrical radius, $v_\mathrm{rot}$ is the circular rotation curve for the galaxy, and $R_0$ is the Galactocentric radial distance to the ``sun'' (taken to be $R_0=8$~kpc). The angular position of the sun was randomly determined, however we found that this did not play a significant role in the distribution of HVC origins (see Appendix~\ref{appendix:rotations}). We calculate the rotation curve using the formula
\begin{equation}
    v_\mathrm{rot}(R) = \sqrt{\frac{GM_\mathrm{enc}(R)}{R^2}}
\end{equation}
where $G$ is Newton's gravitational constant and $M_\mathrm{enc}(R)$ is the total mass enclosed within radius $R$.

Finally, we transform the positions and velocities of all the cold clouds identified above (Section~\ref{sec:cloud_identification}) from Galactocentric coordinates into spherical Galactic coordinates. We then compare the line of sight radial velocity with the calculated $v_\mathrm{disk}$ at that position. Our HVC-analogs are those cold clouds for which the difference is larger than the deviation velocity of 70~\kms\ (as used in \citealt{westmeier18}).

This same methodology is used on each individual gas cell in the simulation in order to reproduce an all-sky map of the high velocity gas. As in the observations \citep{westmeier18}, we calculate the range of possible disk velocities for each line of sight by stepping out in $R$ (assuming the disk extends to cylindrical $R=20$~kpc, and $|z|=5$~kpc). We then exclude all gas particles along that line of sight within 70~\kms\ of this range.
Figure~\ref{fig:onsky} shows the 3D Cartesian representation of this high velocity gas edge-on and face-on in the left two panels. The white outline represents the approximate location of the galactic disk. The top right panel shows how this would appear on the sky in column density from an observer at the solar location within this galaxy. The bottom right panel shows this perspective now colored by line of sight velocity (in the Galactic standard of rest, GSR, frame). The lightness represents column density. This panel also outlines the projected positions of the detected HVC analogs shown in Figure~\ref{fig:cartesian}. We note that there are several large complexes near the equator that are not outlined. This is because they are connected to the galactic disk and are thus excluded from our HVC identification algorithm; however, this gas still appears at anomalous velocities due to the warp of the disk, thus appearing in this figure. \rev{}{There are also several small clouds not identified as HVCs due to their small size ($<10$ gas cells).}

\section{Results} \label{sec:results}

Our fiducial snapshot in which we identify our HVC analogs is snapshot 3230 ($z=0.245$). We find 70 cold clouds with high deviation velocities comprised of more than 10 Voronoi cells below $10^{4.5}$~K. We trace these 70 clouds back in time using the method described in Section~\ref{sec:cloud_tracking} and track their properties in order to determine their origin.

\subsection{Comparison with the MW} \label{sec:mwcomp}

Before tracking our HVCs back to study their origins, we compared the simulated population against current observations of the MW. As stated above, the global properties of the TNG galaxy presented here agree well with those of the MW. The properties of the high-velocity gas also provide good agreement. At its limiting column density ($2.3\times10^{18}$~cm$^{-2}$), the HI4PI survey shows a covering fraction of 11\% for the gas with deviation velocities $>70$~\kms \citep{westmeier18} while our galaxy has a covering fraction of 7.5\% using the same criteria (Figure~\ref{fig:onsky}).

Furthermore, we compare directly to the discrete HVC population identified by \citet{moss13} in the Galactic All-Sky Survey (GASS; \citealt{mcclure-griffiths09}). Restricting to the parameter space accessible by both the survey and our simulation (peak $N_\mathrm{HI}>10^{18}$~cm$^{-2}$ and cloud area $>2$~deg$^{2}$) we find 41 simulated HVCs compared with 30 HVCs in the observational sample (covering the southern sky). And while the median of the peak column density distributions agree well at $10^{19.4}$ and $10^{19.6}$~cm$^{-2}$ for the simulated and observed HVCs, respectively, the areas of the simulated HVCs are generally larger with a distribution median at 35~deg$^{2}$ compared with 6~deg$^2$ for the GASS sample. We believe this is a resolution effect and plan to explore the distribution of HVC projected areas in future works.
These results align with previous works that have generally found good agreement between high velocity gas in the TNG simulations and observations \citep{nelson20,ramesh23}.

\begin{figure*}
    \centering
    \includegraphics[width=\textwidth]{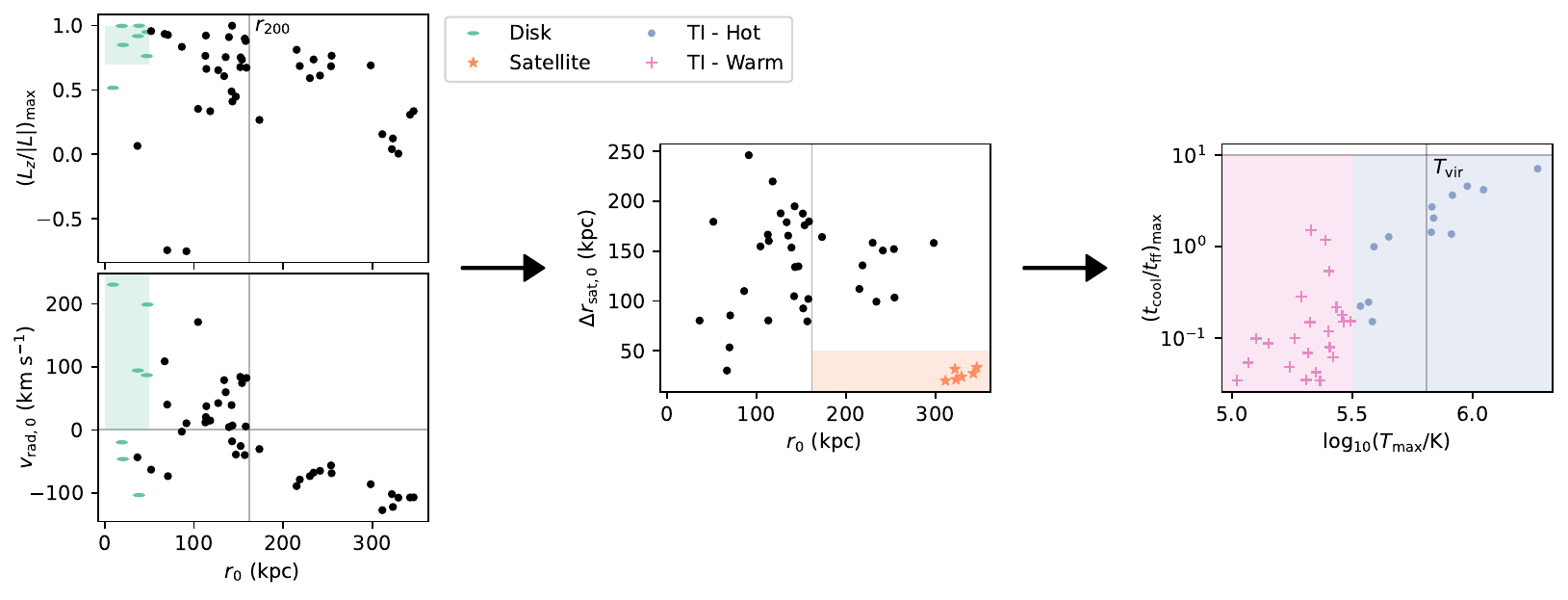}
    \caption{Properties of the HVCs used in classification of their origins. From left to right, we sequentially categorize each HVC based on \rev{four}{six} properties\rev{ plotted with respect to initial 3D radial position of the clouds}{}. First, \rev{}{we define disk origin HVCs as any cloud with high maximum circularity ($(L_z/|L|)_\mathrm{max}$) or positive initial radial velocity ($v_\mathrm{rad,0}$) near the disk.} Any point within the \rev{orange}{green} shaded region is thus classified as a \rev{satellite}{disk} origin HVC. \rev{}{Next,} we use the distance to the closest gas-rich subhalo at $z=0.5$ ($\Delta r_\mathrm{sat,0}$). Finally, the remaining HVCs are classified as forming through thermal instability \rev{}{(TI) by ensuring that the cooling time to freefall time ratios are small enough ($(t_\mathrm{cool}/t_\mathrm{ff})_\mathrm{max}$). They are then subclassified as having formed }\rev{or out of the enriched CGM}{either from the hot or warm CGM} based on a cut in maximum temperature reached over the lifetime of the cloud ($T_\mathrm{max}$). \rev{The bottom right plot shows the maximum value reached for the ratio of the cooling time to the free-fall time ($(t_\mathrm{cool}/t_\mathrm{ff})_\mathrm{max}$) to confirm that the thermally unstable clouds have values less than 10.}{} The points are also colored and shaped based on their determined origin throughout this process -- stars for satellites, ovals for disk, \rev{dots for thermal instability}{and dots and plus symbols for thermal instability from the hot and warm CGM, respectively. Also, once classified, we have hidden the points from subsequent plots for clarity}.}
    \label{fig:classification}
\end{figure*}

\subsection{Identifying Origins} \label{sec:origins}

In order to identify the origins of these clouds, we locate each HVC analog in a six dimensional space of its properties. These individual components are calculated by taking the mean value across all the cloud cells at each snapshot. Several are the initial values corresponding to the property at $z=0.5$ (snapshot 2900), while others are the maximum value attained throughout their history.
\begin{itemize}
    \item $r_0$ -- The initial 3D radius of the cloud cells.
    \item $(L_z/|L|)_\mathrm{max}$ -- The maximum circularity of the cloud in its history. The $z$ direction is aligned with the angular momentum of the disk\rev{}{, thus values close to one mean that the cloud's orbit is more ``disky.''}\footnote{Throughout this paper, we will be considering specific angular momentum, namely $\mathbf{L}=\mathbf{r}\times\mathbf{v}$.}
    \item $v_{\mathrm{rad},0}$ -- The initial radial velocity of the cloud.
    \item $\Delta r_{\mathrm{sat},0}$ -- The minimum separation between the cloud position and all gas-rich satellites (identified using the \texttt{SUBFIND} algorithm) at the initial snapshot.
    \item $T_\mathrm{max}$ -- The maximum temperature reached by the cloud cells in their history.
    \item $(t_\mathrm{cool}/t_\mathrm{ff})_\mathrm{max}$ -- The maximum value of the ratio of the cooling time to the free fall time over the cloud's history.
\end{itemize}

\rev{}{We then sort the HVCs into four different categories based on these properties: disk origin, satellite origin, thermal instability origin from the hot CGM, and thermal instability origin from the warm CGM.}
Satellite origin HVCs must have a small separation from an identified subhalo. Disk origin HVCs exhibit high circularities and positive initial radial velocities. Thermal instability origins \rev{must have reached high maximum temperatures and maintained}{should have} cooling times short relative to their freefall time. While in ideal scenarios this ratio should be less than 1, simulations have shown that cooling can occur for values less than 10 in realistic environments \citep{mccourt12,sharma12}. \rev{The remaining clouds are all relatively metal-enriched at 0.1~$Z_\odot$ and found at all initial radii and radial velocities. We therefore classify these clouds as the ``enriched CGM.''}{The clouds formed through thermal instability are then further classified into hot CGM origin and warm CGM origin based on the maximum temperatures the gas cells reach during their evolution.}

\rev{}{While some of these thermal instability origin clouds reach radii greater than the galaxy's virial radius, they are all relatively metal enriched (at $\sim0.1$~Z$_\odot$, except for a single cloud which starts out at $Z<10^{-3}$~Z$_\odot$). Thus we still classify them as originating from CGM material since they are formed from the metal-enriched gas driven by large scale outflows from the galaxy. We discuss this point further in Appendix~\ref{appendix:metallicity}.}

Our selection criteria are listed below and shown in Figure~\ref{fig:classification} as shaded regions colored by the different origins.\\

\noindent
Disk:
\begin{multline*}
    (r_0 < 50\;\mathrm{kpc})\quad\&\\\Big(((L_z/|L|)_\mathrm{max} > 0.7)\quad|\quad (v_{\mathrm{rad},0} > 0\;\mathrm{km/s})\Big)
\end{multline*}
Satellite:
\begin{equation*}
    (r_0 > r_\mathrm{200})\quad \&\quad (\Delta r_{\mathrm{sat},0} < 50\;\mathrm{kpc})
\end{equation*}
Thermal instability \rev{}{ -- hot CGM}:
\begin{equation*}
    (T_\mathrm{max} > T_\mathrm{vir}/2) \quad\&\quad \Big((t_\mathrm{cool}/t_\mathrm{ff})_\mathrm{max} < 10\Big)
\end{equation*}
\rev{Enriched CGM}{Thermal instability -- warm CGM}:
\begin{equation*}
    (T_\mathrm{max} < T_\mathrm{vir}/2) \quad\&\quad \Big((t_\mathrm{cool}/t_\mathrm{ff})_\mathrm{max} < 10\Big)
\end{equation*}

In addition to using $\Delta r_{\mathrm{sat},0}$ to identify satellite origin HVCs, we isolated the tracer particles associated with each subhalo at $z=0.5$ and propogated them forwards to our final snapshot. We see significant overlap between satellite particles and the largest HVC (located in the southern hemisphere) and nearby clouds. The stellar component of this satellite has not yet merged with the galaxy, lying at a galactocentric radius of 56~kpc. It is spatially separated from the gas by a distance of 59~kpc due to ram pressure and drag forces. The stellar component is still colocated with an identified \texttt{SUBFIND} subhalo.

Interestingly, we also see two other satellites in different stages of disruption. For the first, both the stars and the gas have been integrated into the galactic disk. For the second, the gas content of the satellite now resides within the central galaxy's disk, however the stellar component (and DM halo) is still undergoing tidal disruption and tidal tails can clearly be seen in the distribution of stars.

\begin{figure}
    \centering
    \includegraphics[width=\columnwidth]{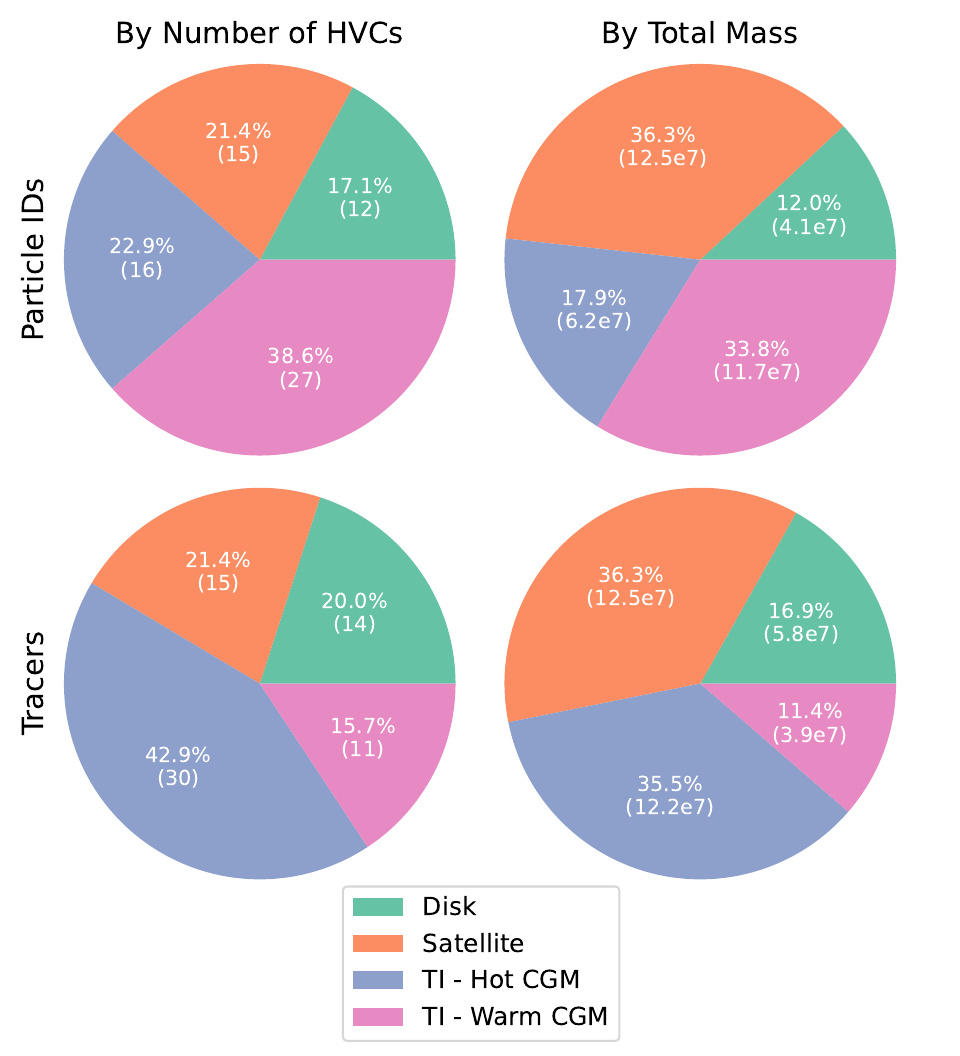}
    \caption{The distribution of HVC origins. On the left we count up the number of individual clouds each coming from \rev{the disk, thermal instability, the IGM, and satellite accretion}{the galactic disk, stripped satellite material, and thermal instability from the hot and warm CGM}. While on the right we sum up the total mass of all the clouds in each category. In the top plots, the origins were determined based on properties of the tracked particle IDs. While in the bottom plots, we used the properties of the tracer particle locations.}
    \label{fig:pies}
\end{figure}

\subsection{Distribution of Origins} \label{sec:origindist}

Tracking gas cell IDs, we find 12 HVCs originating from the galactic disk (12\% by mass), 15 originating from stripped satellite material (36\% by mass), \rev{16}{and 43} forming through thermal instability (\rev{18}{52}\% by mass), \rev{and 27 that are part of the enriched CGM (34\% by mass)}{with 16 coming from the hot CGM (18\% by mass) and 27 coming from the warm CGM (34\% by mass)}. These categories are discussed further below, and this distribution is shown in the top panels of Figure~\ref{fig:pies}.

Figure~\ref{fig:pies} also shows the distribution of origins using properties of the tracer particles instead of the gas cell IDs in the bottom panels. Using tracer particles we find many more clouds \rev{classified as formed via thermal instability}{originating from the hot CGM} because the maximum temperatures reached by these particles are on average higher due to their increased diffusivity. Since the hot phase of the CGM is the volume-filling component, the tracers are more likely to diffuse into this phase while the gas cells themselves are more likely to remain in the dense, mass-dominant phase. This change has only a very minor effect on the disk, satellite\rev{}{, and IGM} origin HVCs. The specific cuts and results for the tracer analysis are discussed in Appendix~\ref{appendix:tr}.

\rev{}{We can estimate the ``errors'' on these distributions a variety of ways. First of all, by taking different solar positions within our galaxy (as discussed in Appendix~\ref{appendix:rotations}), we can understand how these origins vary based on observational effects such as the definition of the deviation velocity. The mean and standard deviation of the origin percentages from these four realizations are $18.9\pm 3.9\%$, $17.2\pm 2.9\%$, $24.3\pm 1.6\%$, and $39.7\pm 1.8\%$ for the disk, satellite, hot CGM thermal instability, and warm CGM thermal instability, respectively. Furthermore, we can estimate the galaxy-to-galaxy variance by attributing a Poisson error to each origin pathway due to the small number of HVCs. For our fiducial origin distribution (top left of Figure~\ref{fig:pies}), our errors are 4.9\%, 5.5\%, 5.7\%, and 7.4\% again for the disk, satellite, hot CGM, and warm CGM, respectively.}

\begin{figure*}
    \centering
    \includegraphics[width=\textwidth]{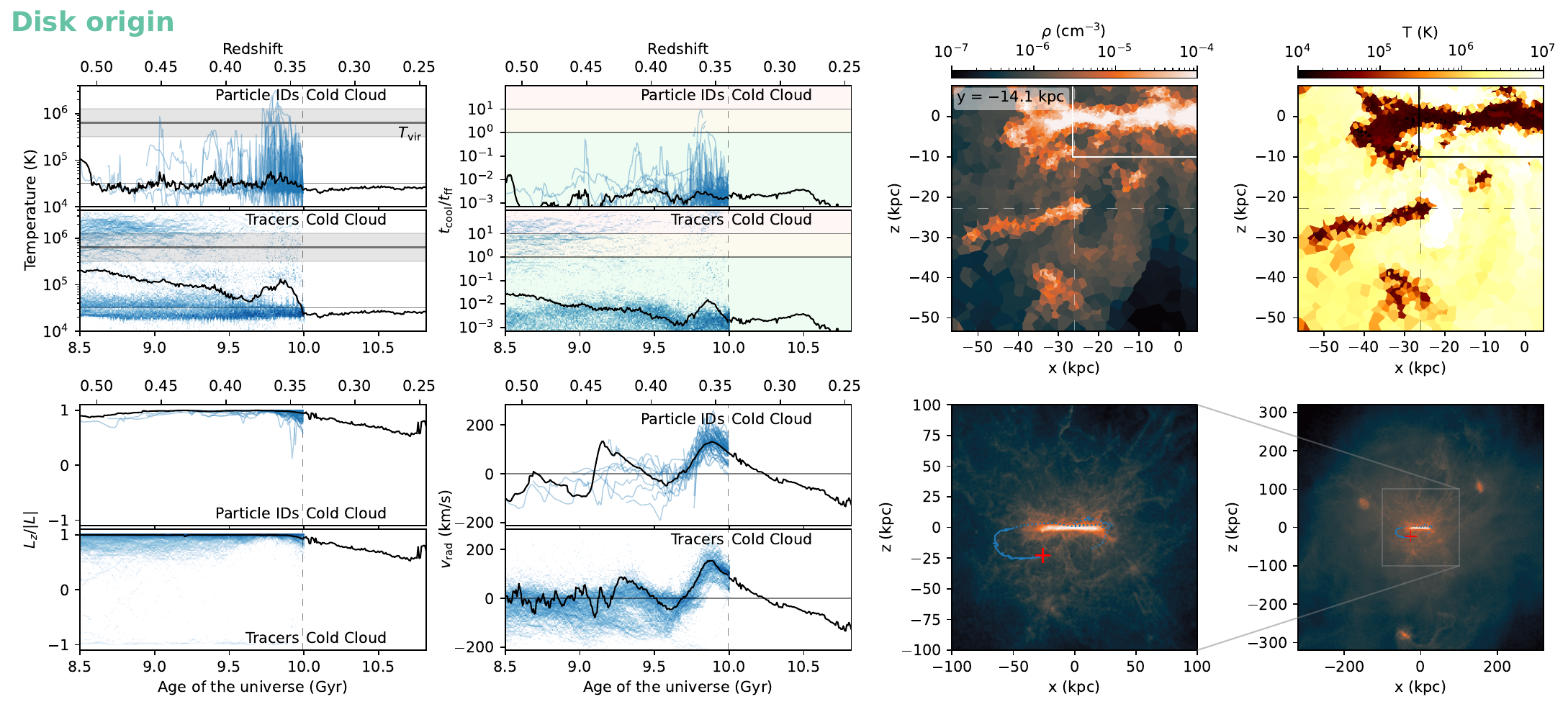}
    \includegraphics[width=\textwidth]{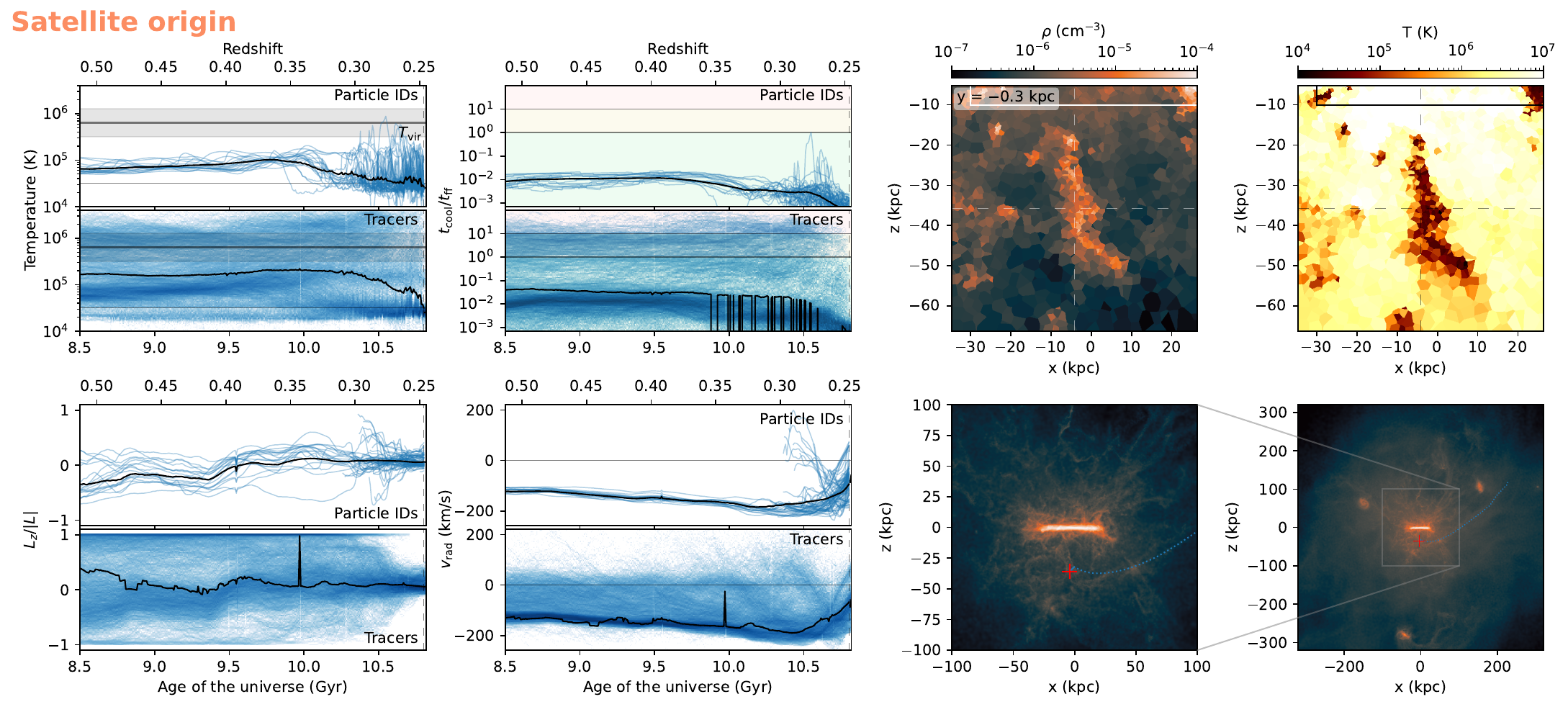}
    \caption{Properties of example clouds originating from the disk and a satellite. For each example cloud, we show four panels on the left depicting the history of the HVC analog and its precursor particles or tracers. The top half of each panel shows the history by tracking particle IDs, while the bottom half shows the history of the tracer particles. At the dashed vertical line, the cloud is identified as a contiguous group of cold Voronoi cells and is tracked using the method described in Section~\ref{sec:cloud_tracking}. To the left of the dashed vertical line, the blue distribution shows the history of each cell or tracer particle while the black line denotes the mean of all the cells. From the top left, clockwise, the panels depict temperature, the ratio of $t_\mathrm{cool}$ to $t_\mathrm{ff}$, the 3D radial velocity, and the circularity ($L_z/|L|$). The four panels on the right show the appearance of the cloud at the final snapshot ($z\sim0.25$). The top two panels show a slice through the domain centered on the HVC analog and show 3D mass density on the left, and temperature on the right. An approximate demarkation of the galactic disk is shown as a rectangle (if applicable). The bottom two panels show projected gas density in the background with an overlay of the HVC trajectory shown in blue (the dotted line denotes when it is being tracked by particle IDs, while the solid line is once it is identified as a cold cloud) and the cloud's final position as a red plus symbol.}
    \label{fig:combined1}
\end{figure*}

\begin{figure*}
    \centering
    \includegraphics[width=\textwidth]{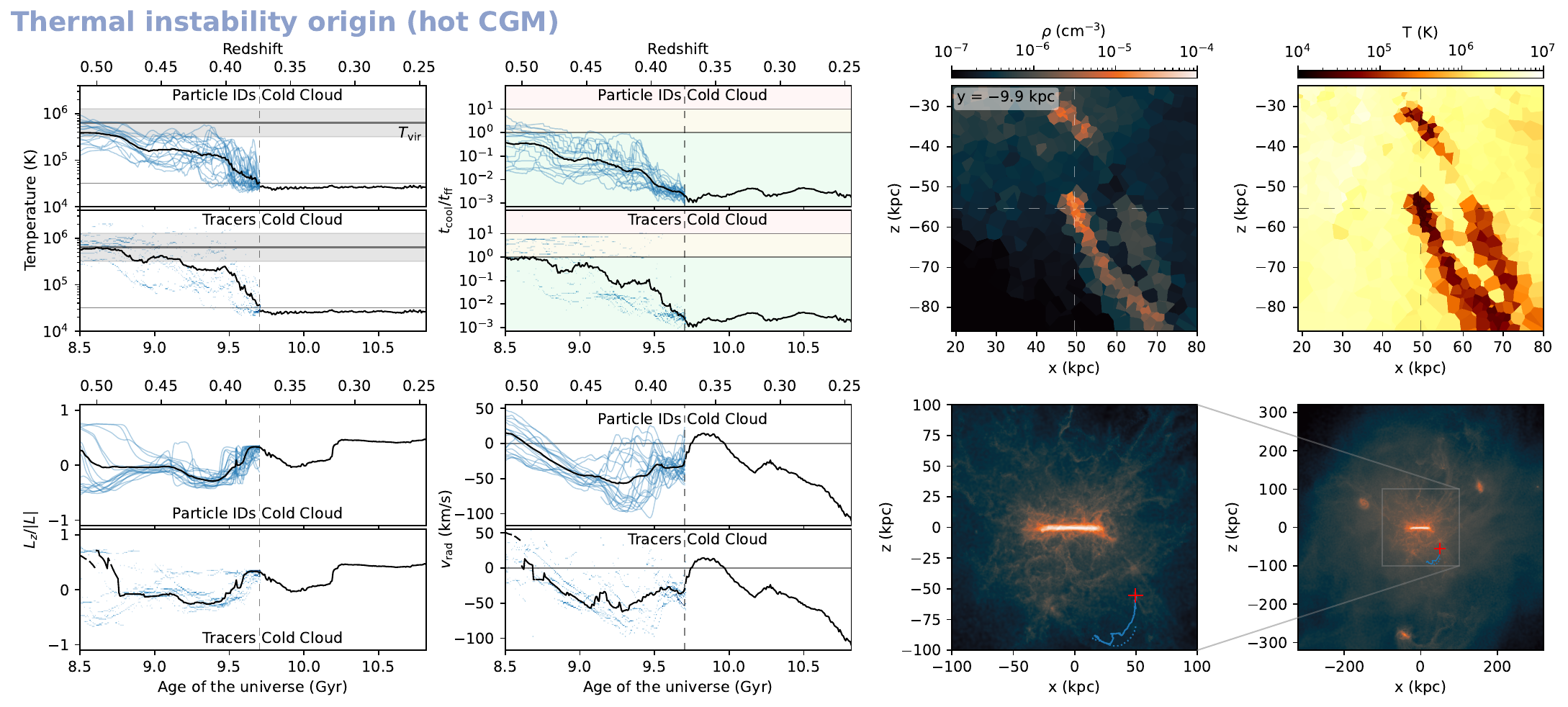}
    \includegraphics[width=\textwidth]{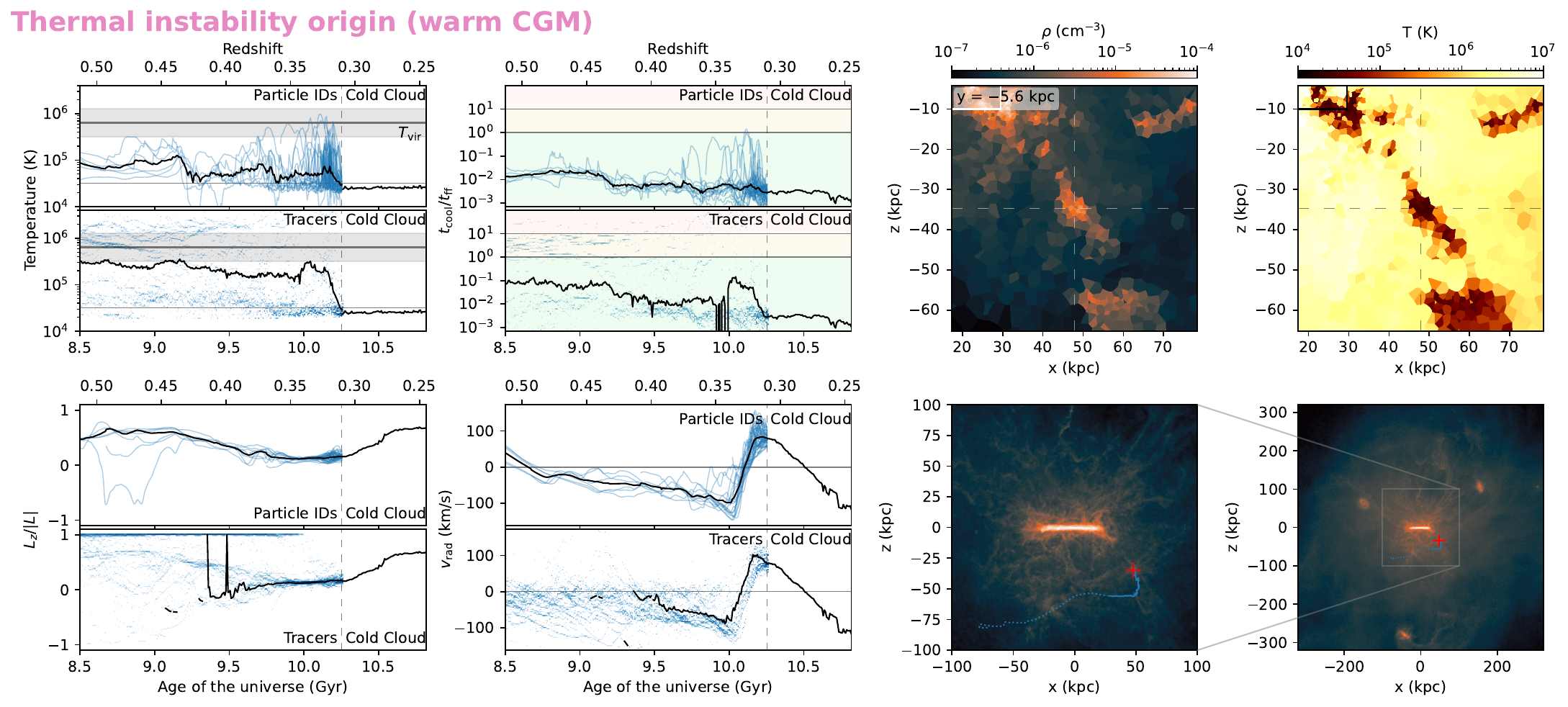}
    \caption{Continuation of Figure~\ref{fig:combined1} showing the properties of example clouds \rev{originating}{forming} via thermal instability \rev{and the enriched CGM}{from the hot and warm CGM}.}
    \label{fig:combined2}
\end{figure*}

\subsection{Cloud Histories} \label{sec:histories}

We are also able to track the individual HVCs and their precursors throughout cosmic time. Figures~\ref{fig:combined1} and \ref{fig:combined2} show four example clouds, one for each origin pathway. The panels on the left show the evolution of the cloud's temperature, cooling time ($t_\mathrm{cool}/t_\mathrm{ff}$), angular momentum ($L_z/|L|$), and radial velocity. The black lines on the right sides of these plots (in the region labelled ``Cold Cloud'') track the mean properties of the identified HVC as a collection of Voronoi cells with $T<10^{4.5}$ K.

However, as we move backward in time, at some point the cloud heats up and we can no longer track it this way. Here we show two different methods of tracking the HVC's precursor gas in the two sub-panels in each plot. On the top we show the evolution of the specific gas cells that comprise the HVC at first identification (using the cell IDs). The black line is the mean value and the blue lines are the individual gas cells. As discussed in Section~\ref{sec:cloud_tracking}, this method does not fully capture the gas flows between cells, however it provides a limit in the minimum diffusion case.

\begin{figure*}
    \centering
    \includegraphics[width=0.9\textwidth]{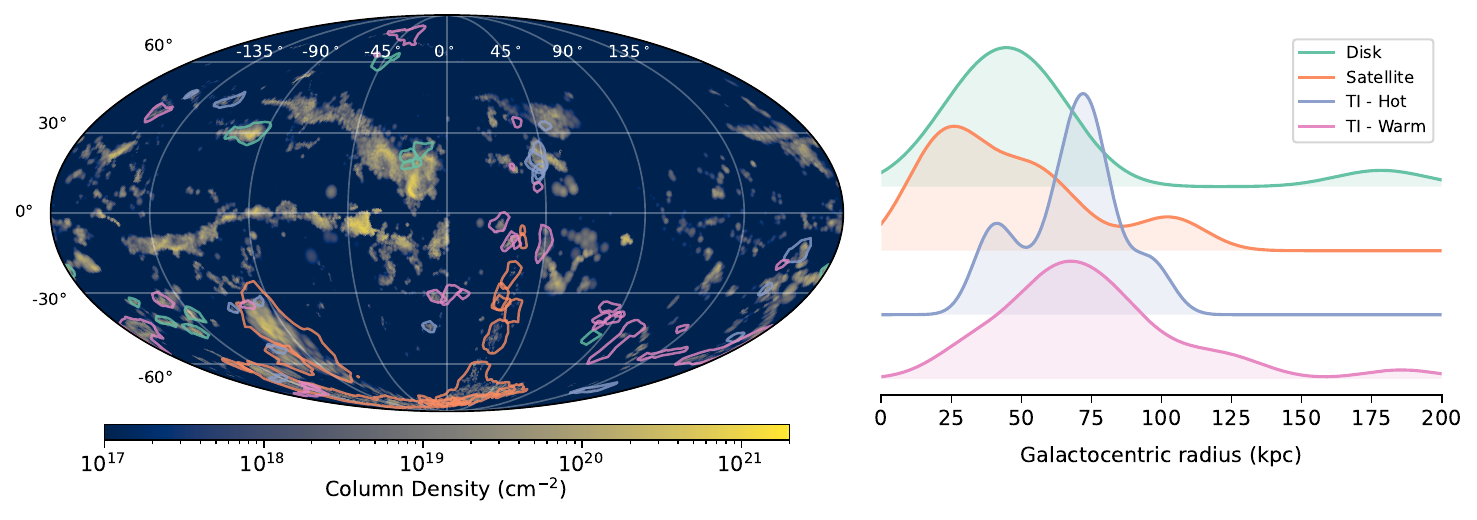}
    \caption{The on-sky and radial distributions of the HVCs by origin. The left panel shows all the high-velocity gas (as in Figure~\ref{fig:onsky}) with each of the HVCs outlined in a color representing their origin. The right panel shows the distribution of HVCs as a function of Galactocentric radius (with each curve offset in y for clarity).}
    \label{fig:onsky_origin}
\end{figure*}

In the bottom sub-panel we show the evolution of the gas tracer particles that were in the HVC at identification. Again, the black line is the mean, but the blue background depicts a histogram showing the distribution of tracer properties (since there were so many tracers that individual lines became unreadable). As discussed in \citet{genel13}, the tracers diffuse more quickly than the gas and thus this panel gives us a limit in the maximum diffusion case. By comparing these two methods, we can be sure that our analysis is consistent and robust.

The four panels on the right depict trajectory of the HVC as well as its final position and appearance. The top panels show the density and temperature of the HVC (located in the center of each plot) with the galactic disk shown outlined in a white rectangle. The bottom panels show two zoomed out views of the galaxy with the HVC's trajectory shown in blue and its final position shown as a red plus. The blue dashed line is the mean trajectory of the precursor traced by gas cell IDs, while the solid blue line is the mean trajectory of the tracked cold cloud.

For the disk origin example (Figure~\ref{fig:combined1}, top), we can see that it remains cold throughout its history and has a circularity of $\approx$1 for the first $\sim$half of its evolution. For the satellite origin HVC (Figure~\ref{fig:combined1}, bottom), it also remains relatively cool however it has a consistent angular momentum and radial velocity as it approaches the galaxy from large distances. For the example cloud of thermal instability origin \rev{}{from the hot CGM} (Figure~\ref{fig:combined2}, top), we see a steady decrease in temperature until the cloud forms with $t_\mathrm{cool}/t_\mathrm{ff}$ remaining below 1 throughout. \rev{Finally, the enriched CGM origin (Figure~\ref{fig:combined2}, bottom) shows a relatively constant, cool temperature with slowly varying radial velocity and smooth trajectory. This category has several different populations of clouds with a variety of initial radii and radial velocities. However, they all have cool max temperatures and metallicities $\sim0.1$ Z$_\odot$.}{Finally, the warm CGM thermal instability origin (Figure~\ref{fig:combined2}, bottom) shows a similar variation in radial velocity to the hot CGM origin, indicating that the dynamics of these cloud are affected by the turbulent flows within the CGM. However, its temperature remains relatively low throughout its history and undergoes several bouts of rapid cooling before forming a surviving cold cloud.}

\begin{figure*}
    \centering
    \includegraphics[width=0.9\textwidth]{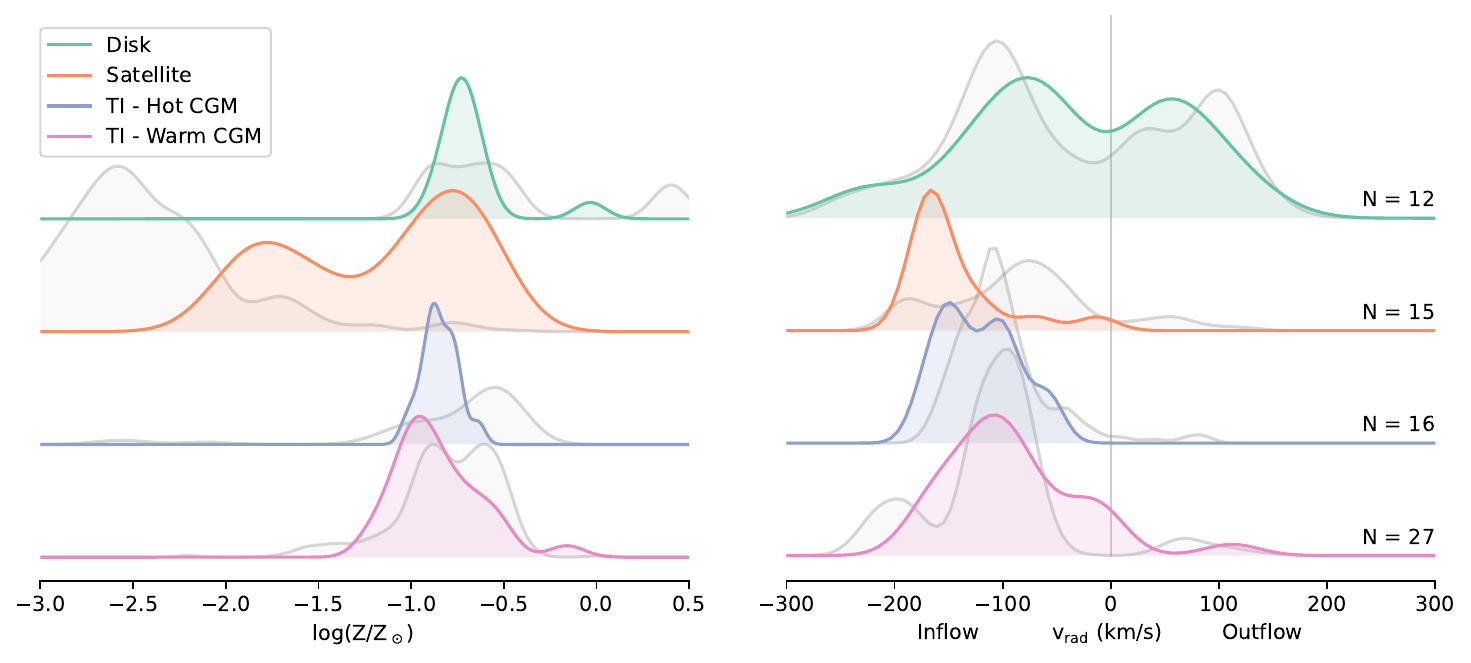}
    \caption{The final snapshot properties of the HVCs separated by origin. The left four panels show the final snapshot metallicity distribution in color and \rev{the initial distribution of the cloud progenitor cells in grey}{in grey we show the distribution of metallicities for all of the clouds' progenitor gas cells at $z=0.5$ (the ``initial distribution'')}. \rev{The vertical lines denote the mean metallicity of the disk, CGM, IGM, and satellite galaxies for reference.}{} On the right we show the distribution of 3D radial velocities with positive velocities denoting outflow. As expected, disk origin HVCs are the only outflowing clouds.}
    \label{fig:hvc_dist}
\end{figure*}

\subsection{Observable Properties} \label{sec:obsvprops}

Figures~\ref{fig:onsky_origin} and \ref{fig:hvc_dist} show the distributions and properties of the HVC sample by origin. Figure~\ref{fig:onsky_origin} shows the onsky (left) and radial (right) distributions for each of the different formation pathways. Most of the satellite origin clouds are in the southern hemisphere close to the location of the infalling satellite. And the inner-most clouds are generally from satellites or the disk.

Figure~\ref{fig:hvc_dist} shows the distributions of metallicities (left) and galactocentric radial velocities (right) for our HVC sample. Each panel shows the distribution for a different origin (classified as shown in Figure~\ref{fig:classification}). In the left panels, \rev{the vertical dashed lines denote the mean metallicities of the source gas (disk, satellite ISM, and circumgalactic gas).
Additionally,}{} the grey background shows the initial distribution of the cells comprising the HVCs at $z=0.5$. We see that most of the HVCs have metallicities of $Z\sim0.1$ Z$_\odot$, and are infalling towards the galaxy. We discuss the implications of these results for observations below.

\section{Discussion} \label{sec:discussion}

While we have focused on the origin of HVC analogs, there have been several other works exploring the origin of the CGM as a whole. \citet{hafen19} use the FIRE simulations \citep{fire1,fire2} to study the CGM of MW-mass halos. For $z\sim0.25$, they find 16\% of the CGM comes from satellites, 20\% from recycling of the galaxy's own winds, and 53\% originates in the IGM. \citet{decataldo24} also identify the origins of the cold CGM using the Eris2K simulations. For $z\sim0.5$, they find 1.5\% entering the CGM from the IGM in the cold phase, 2.5\% from satellites, 10\% from the galactic disk, and 84\% cooled from hot gas while bound to the galaxy. \rev{}{Using the GIBLE simulations at $z=0$ \citep{ramesh24a}, \citet{ramesh24b} find a distribution of cold clouds equally split between disk outflows and cooling from the CGM ($\sim45\%$ each), with minor contributions from satellites ($\lesssim5\%$).}

These classifications vary slightly across the literature, but there are a few key differences with our results here. Notably the HVC analog mass fraction originating from satellites is 36\% compared with 2.5\% of the cold CGM in general \citep{decataldo24} and \rev{}{$5-16$\%} of the CGM in total \citep{hafen19,ramesh24b}. This shows that stripped material from satellites is more likely to appear at high velocities relative to the galactic disk rotation than thermally unstable CGM gas or galactic outflows.

More consistent with the global CGM is the distribution of the remaining material. Roughly 15\% of the HVC analogs (by mass) originate in the disk, while the remaining material cools in the CGM ($\sim$50\%). This is comparable to the previous results listed above. \rev{}{\citet{ramesh24b} find a more significant contribution from disk material, however a full investigation of the role played by subgrid feedback models is important to understand this component.}

It has also been shown that the gaseous histories of galaxies depend greatly on numerical effects \citep[e.g.][]{nelson13,hobbs13}. Early smoothed-particle hydrodynamics (SPH) codes were unable to accurately model various instabilities which led to the increased stability of cold gaseous filaments in the CGM. Using more modern simulation techniques, we have seen that the fragmentation of circumgalactic gas is ubiquitous with resolution and numerical methods playing an important role in the realism of the results \citep{vandevoort19,foggie,suresh19,gible}.

\rev{}{Additionally, there have been several works comparing simulated HVCs against the MW population. \citet{connors06} use early SPH cosmological zoom simulations in a $\Lambda$CDM universe. They find anomalous velocity \ion{H}{1} clouds with similar numbers and column densities to those found around our Galaxy. \citet{ramesh23} perform a more in-depth analysis with the TNG50 MW/M31 sample \citep{pillepich23} also confirming that the HVC population found in simulated MW-like galaxies is in good agreement with the data, as we also find in this work.}

On smaller scales, wind tunnel simulations have shown that cold clouds moving through a hot gaseous medium can survive and grow (depending on the initial properties; \citealt{kwak11,gronke18,gronke22}). \rev{The criteria for cloud survival is often given in terms of critical size; large clouds will grow, and small clouds will shatter. Pulling values from our simulation, we see that our clouds have overdensities of $4-10$ and Mach numbers of $0.2-0.9$. Using equation (2) from \citet{gronke22}, we find critical radii on the order of a few parsecs. Due to the resolution of the simulation, all our clouds have sizes $>1$~kpc, so we would expect them to grow. They}{Furthermore, more realistic simulations of cloud evolution in a stratified medium have shown that this is still possible, however the criteria are somewhat more stringent \citep{tan23}. Pulling values from our simulation, we see that our clouds have overdensities of $4-10$, Mach numbers of $0.2-0.9$, and sizes of $1-3$~kpc. Using equations (11), (24), and (27) from \citet{tan23}, we find ratios of $t_\mathrm{grow}/t_\mathrm{cc}$ ranging from 0.8-1.7 which is less than $f_S=4$, so our HVC analogs should survive. \citet{tan21}} also specify a resolution requirement to resolve the cloud by at least $\sim$~16~cells, and our clouds contain $13-137$ cells at identification which should be sufficient in most cases.

Moreover, we can compare the simulation resolution to an estimate of the fragmentation length to confirm that we are able to resolve the thermal instability. \citet{maller04} derive the size of clouds formed in a two-phase pressure confinement model of thermal instability driven fragmentation. Using their equation (26), we can calculate the expected sizes of clouds that will form in the CGM of our galaxy. By using the mass of cold connected cells and the temperature of the surrounding ambient gas, we find cloud sizes of $1-5$~kpc which is in agreement with their actual sizes ($r_\mathrm{HVC}/\ell_\mathrm{frag}\sim0.5-1.75$). This reassures us that at the resolution of TNG50-1, we will be able to see clouds condense due to thermal instability in the CGM.

Finally, we return to the observability of these different origin pathways. As shown in Figure~\ref{fig:hvc_dist}, we see that most of the present-day HVCs have metallicities of $\sim0.1$ Z$_\odot$. \rev{presumably due to mixing}{This is due to strong gas mixing which has been shown to play a significant role in the evolution of HVCs on small scales \citep{heitsch22}}. However, we do see a few lower metallicity clouds originating from \rev{satellite accretion}{the material stripped out of satellites}. While we do not know the origins of many of the MW HVCs, this does seem to be in agreement based on the metallicity observations that we do have.
\rev{The Smith Cloud and Complex C have metallicities of $0.3-0.5$ Z$_\odot$ \citep{fox16,fox23} while the Magellanic Stream and Leading Arm (material tidally stripped from the Large and Small Magellanic Clouds) have lower metallicities of $\sim0.1$ Z$_\odot$ \citep{fox18}.}{
The Smith Cloud, Complex K, Cloud MIII, and the PP Arch all have metallicities at or above 0.3~Z$_\odot$, and Complex WE, the IV Arch, and the LLIV Arch have supersolar metallicities \citep{hayakawa24}. On the other hand, the Magellanic Stream and Leading Arm (materially tidally stripped from the Large and Small Magellanic Clouds) have lower metallicities of $\sim0.1$~Z$_\odot$ (with a few isolated sightlines up to 0.3 and 0.5 Z$_\odot$; \citealt{fox18}). However, there are also a few other HVCs with low metallicities (Complex A at 0.06 Z$_\odot$, and Complex C at $0.1-0.3$ Z$_\odot$; \citealt{hayakawa24}).}

\rev{}{There have also been several recently identified compact high-velocity clouds (CHVCs) with very low metallicities of a few percent solar \citep{ashley24}. Consistent with their conclusions, our work would predict that these CHVCs originated from external satellites due to their low metallicities. While we are unable to resolve CHVCs in most cases with the TNG50 simulation (see also Section~\ref{sec:mwcomp}), future work with enhanced resolution will allow us to explore the origins of this population directly.}

In the right panels of Figure~\ref{fig:hvc_dist}, we see that nearly all the simulated HVCs are infalling (negative $v_\mathrm{rad}$). The only formation pathway that leads to significant outflowing gas is disk origin, as expected. In our observational sample of HVCs, we also see the majority of the gas infalling towards us, however transforming from Heliocentric coordinates to Galactocentric coordinates without proper motions for these gas clouds can be imprecise \citep{wakker97}. Most of the positive velocity clouds are very small, which will allow them to be more easily pushed and pulled by turbulence in the inner CGM.

\subsection{Further Considerations} \label{sec:caveats}

Gas mixing plays a key role in these processes and as stated above, has a strong dependence on numerical methodologies \citep{springel10,vogelsberger12,sijacki12,hayward14}. It has been shown that SPH codes are unable to produce certain instabilities which results in a lack of mixing and cooling while adaptive-mesh refinement (AMR) and Eulerian codes suffer from overcooling and preferred directions. \arepo's moving-mesh scheme based on Voronoi tessellation mitigates these issues and allows for appropriate gas mixing and cooling. While a full comparison between cosmological scales and those of the turbulent mixing layers governing gas mixing should be performed \citep{fielding20,heitsch22}, \arepo's numerical model is theoretically able to capture instabilities and hydrodynamic mixing correctly. However, resolution of the simulations plays an important role in this issue as well.

When considering simulations, especially studying the CGM, resolution and computational feasability are continuously at odds. IllustrisTNG was designed to be a set of cosmological box simulations, so while they are cutting edge, the baryonic resolution is limited to $4.5\times10^5$~\msol. However, as stated above, the spatial resolution is sufficient to resolve the fragmentation length for thermal instabilities \citep{maller04}. This is consistent with previous works that have studied the thermally unstable CGM in TNG50 \citep{nelson20}. A full convergence study is required and we will cover this in future work utilizing cosmological zoom-in simulations with higher resolution.

\rev{}{Finally, this work includes the study of a single galaxy. Galaxies are complex, and minor differences in their properties can lead to dramatic differences in CGM structure. A study expanding this work to other simulated galaxies would also be important to learn about the range in possible formation pathways of HVCs even within the context of our own Milky Way.}

\section{Conclusions} \label{sec:conclusions}

In this paper, we have identified 70 \rev{}{\ion{H}{1}} HVC analogs in a TNG50-1 galaxy. Using the high time cadence of \texttt{subbox0}, we are able to track these HVCs in time to determine their origins.
\begin{itemize}
    \item We find a realistic distribution of high velocity gas which is well captured by our algorithm.
    \item By identifying cold clouds at each snapshot and propagating them forward in time, we can track the transient HVC as gas cycles in and out of it.
    \item For the first time, we are able to identify HVC analogs formed via thermal instability. Due to the intense turbulence in the CGM, these clouds can be pushed to high deviation velocities consistent with the observed HVCs in the MW. \rev{23}{62\%} of our HVC analogs are of thermal instability origin \rev{}{with 23\% cooling from hot material in the CGM and 39\% formed out of the CGM's warm phase}.
    \item We also find 17\% originating from the disk \rev{,}{and} 21\% coming from satellites\rev{, and 39\% comprised of relatively high metallicity circumgalactic gas at a variety of radii (the enriched CGM)}{}.
    \item We obtain similar results by using both gas cell IDs and Monte Carlo tracer particles to track the HVC analogs back before they were cold.
    \item Based on this analysis, low metallicity HVCs are most likely to be associated with a satellite infall, and positive radial velocity clouds are most likely of disk origin.
\end{itemize}

As we increase the resolution in the CGM, more and more structure continues to appear. By applying this technique to new, cosmological zoom and CGM-refinement simulations, we will be able to get a better understanding of the low-mass end of the HVC population. Moreover, we will be able to explore the future of these clouds as they interact with the galactic disk and facilitate star formation and galaxy growth.

\begin{acknowledgments}

The authors thank the anonymous referee for their constructive comments.
The authors thank Sarah Jeffreson, Jake Bennett, and Cassi Lochhaas for useful discussions in the development of this paper. SL thanks Andrew J. Fox for his expertise on this topic and his insightful comments on the manuscript. The computations in this paper were run on the FASRC cluster supported by the FAS Division of Science Research Computing Group at Harvard University.

\end{acknowledgments}

\appendix

\section{Azimuthal Solar Position}
\label{appendix:rotations}

Our galaxy was initially rotated from its position within the TNG box such that the disk's angular momentum was aligned with the $z$-axis. Specifically, it was rotated by the angle $\cos^{-1}(\hat{L}\cdot \hat{z})$ around the vector $\hat{L}\times\hat{z}$. The fiducial azimuthal position of the ``Sun'' is therefore $(x,y,z)=(-8,0,0)$~kpc in this rotated coordinate system.

\begin{figure}[h!]
    \centering
    \includegraphics[width=\textwidth]{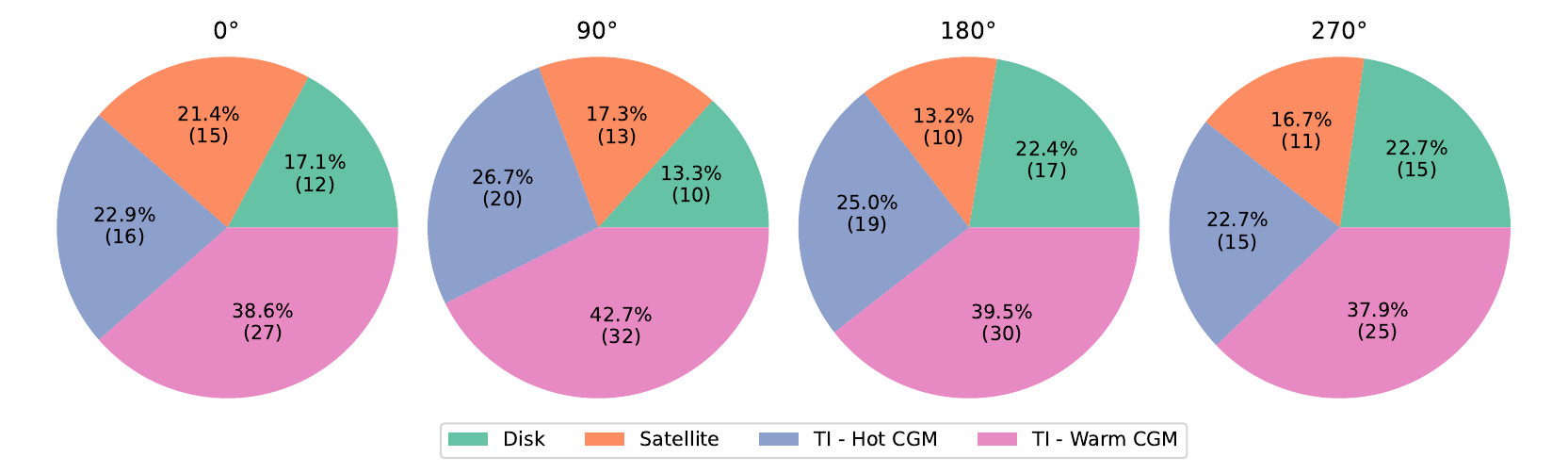}
    \caption{Similar to Figure~\ref{fig:pies} showing the number of identified HVCs by origin for different solar positions. The left-most panel matches that in Figure~\ref{fig:pies} and the subsequent three panels show the distributions after rotating the solar position by 90, 180, and 270 degrees.}
    \label{fig:angles}
\end{figure}

We also performed the full analysis after rotating the galaxy by 90, 180, and 270 degrees to determine the impact, if any, of the azimuthal solar position on our results. Using the full set of 211 identified cold clouds at $z=0.25$, we calculated the deviation velocities for each of these different viewing angles. We found the number of clouds identified as ``high velocity'' ranged from 68 to 78 with 41 consistent across all four angles. Furthermore, Figure~\ref{fig:angles} shows the distribution of origins for each of these solar positions. While there is some slight variation, the overall relative distributions are consistent.

\section{Tracer Particle Results}
\label{appendix:tr}

\begin{figure*}[h!]
    \centering
    \includegraphics[width=0.9\textwidth]{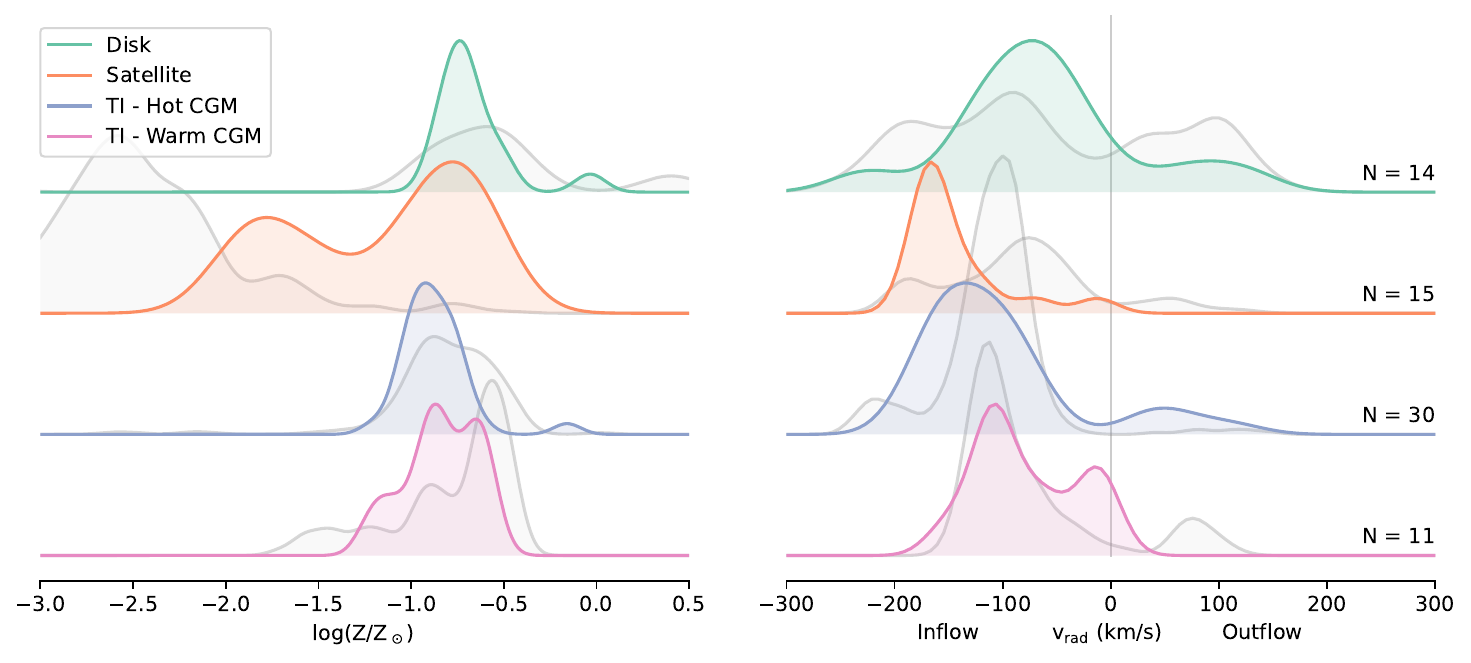}
    \caption{As Figure~\ref{fig:hvc_dist} except with the origins determined using the tracer particle values. This leads to an increase in the number of disk origin clouds and thermal instability clouds, while decreasing the number of IGM clouds. However, the overall trends are unchanged.}
    \label{fig:hvc_dist_tr}
\end{figure*}

Figure~\ref{fig:pies} shows the distribution of origins determined for both particle ID tracking and tracer particle tracking. However, due to the inflated temperatures seen by the tracer particles, a few modifications to the selection criteria were required. The cutoff temperature between thermal instability and the enriched CGM origins was increased from $T_\mathrm{vir}/2$ to $T_\mathrm{vir}$, and the minimum separation to a satellite, $\Delta r_{\mathrm{sat},0}$, was also increased from 50 to 60~kpc. The resulting property distributions are shown in Figure~\ref{fig:hvc_dist_tr}.

\section{Metallicity of the IGM} \label{appendix:metallicity}

Interestingly, we did not find any HVCs originating from very low metallicity IGM gas. All of the clouds that form at large radii have relatively high metallicities of $\sim0.1$ Z$_\odot$. This seems to be due to very strong outflows coming from the galaxy's supermassive black hole which is propelling high metallicity material out to hundreds of kpc away from the galaxy \citep{suresh15}. These outflows prevent IGM gas from falling directly into the galaxy's CGM, and mixes strongly with the low and intermediate metallicity material. Additionally, small-scale simulations show that high velocity gas experiences very strong mixing and can raise or lower a cloud's metallicity dramatically within a few hundred Myrs \citep{gritton14,heitsch22}.

\begin{figure}[h!]
    \centering
    \includegraphics[width=0.75\textwidth]{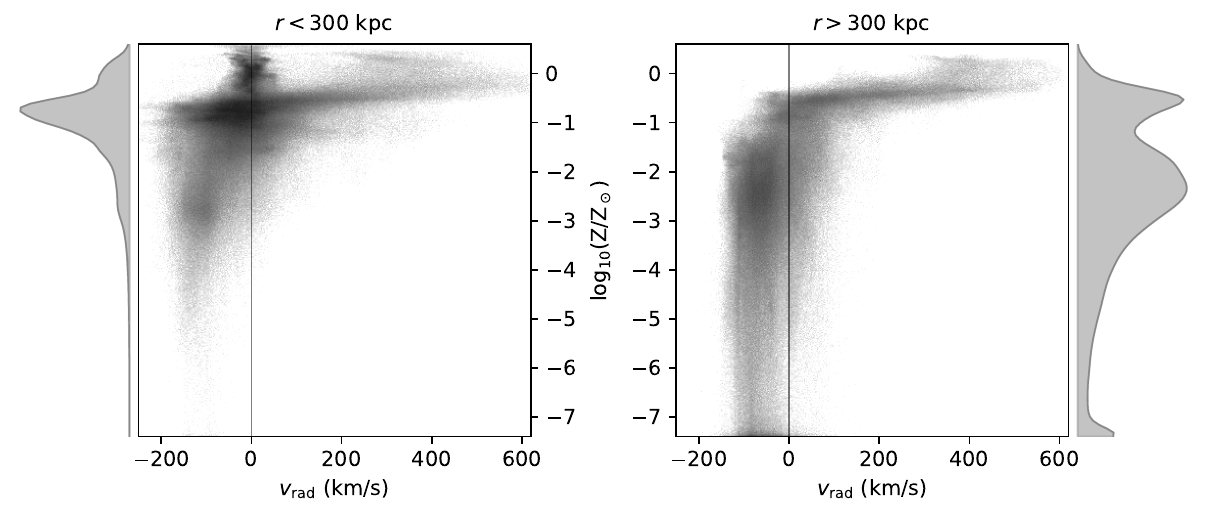}
    \caption{The metallicity distribution of gas within and beyond 300~kpc as a function of radial velocity. On the left we show $r<300$~kpc gas, and on the right $r>300$~kpc gas. The shaded curves on the far left and right are the total metallicity histograms for the $r<300$~kpc and $r>300$~kpc gas, respectively. A vertical line denotes zero velocity with inflows as negative and outflows as positive.}
    \label{fig:igmz}
\end{figure}

Figure~\ref{fig:igmz} shows the distribution of gas within (left) and beyond (right) 300~kpc in metallicity vs radial velocity. Inflows correspond to negative $v_\mathrm{rad}$ while outflows are positive. On the left, we can see inflowing low metallicity material and outflowing high metallicity material. On the right, we see a similar trend even at these very large radii. There are metal enriched outflows at the level of $>0.1$~Z$_\odot$ in addition to low metallicity infalling material. The integrated histograms for these two radial cuts are shown on the left and right (for the $r<300$~kpc gas and the $r>300$~kpc gas, respectively). Interestingly, the very low metallicity gas ($Z<0.01$~Z$_\odot$) does not survive within 300~kpc. This is consistent with our findings that the HVC analogs that we detect originating at large radii have $Z\sim0.1$~Z$_\odot$ indicating that they were formed out of this metal rich outflow rather than the pristine intergalactic medium.

\newpage
\bibliography{references}{}

\begin{thebibliography}{}
\expandafter\ifx\csname natexlab\endcsname\relax\def\natexlab#1{#1}\fi
\providecommand{\url}[1]{\href{#1}{#1}}
\providecommand{\dodoi}[1]{doi:~\href{http://doi.org/#1}{\nolinkurl{#1}}}
\providecommand{\doeprint}[1]{\href{http://ascl.net/#1}{\nolinkurl{http://ascl.net/#1}}}
\providecommand{\doarXiv}[1]{\href{https://arxiv.org/abs/#1}{\nolinkurl{https://arxiv.org/abs/#1}}}

\bibitem[{{Ashley} {et~al.}(2024){Ashley}, {Fox}, {Lockman}, {Wakker},
  {Richter}, {French}, {Moss}, \& {McClure-Griffiths}}]{ashley24}
{Ashley}, T., {Fox}, A.~J., {Lockman}, F.~J., {et~al.} 2024, \apj, 961, 94,
  \dodoi{10.3847/1538-4357/ad0cb7}

\bibitem[{{Binney}(1977)}]{binney77}
{Binney}, J. 1977, \apj, 215, 483, \dodoi{10.1086/155378}

\bibitem[{{Bland-Hawthorn} \& {Gerhard}(2016)}]{bland-hawthorn16}
{Bland-Hawthorn}, J., \& {Gerhard}, O. 2016, \araa, 54, 529,
  \dodoi{10.1146/annurev-astro-081915-023441}

\bibitem[{{Bland-Hawthorn} \& {Maloney}(1999)}]{bland-hawthorn99}
{Bland-Hawthorn}, J., \& {Maloney}, P.~R. 1999, \apjl, 510, L33,
  \dodoi{10.1086/311797}

\bibitem[{{Bregman}(1980)}]{bregman80}
{Bregman}, J.~N. 1980, \apj, 236, 577, \dodoi{10.1086/157776}

\bibitem[{{Chevalier} \& {Oegerle}(1979)}]{chevalier79}
{Chevalier}, R.~A., \& {Oegerle}, W.~R. 1979, \apj, 227, 398,
  \dodoi{10.1086/156744}

\bibitem[{{Connors} {et~al.}(2006){Connors}, {Kawata}, {Bailin}, {Tumlinson},
  \& {Gibson}}]{connors06}
{Connors}, T.~W., {Kawata}, D., {Bailin}, J., {Tumlinson}, J., \& {Gibson},
  B.~K. 2006, \apjl, 646, L53, \dodoi{10.1086/506519}

\bibitem[{{Decataldo} {et~al.}(2024){Decataldo}, {Shen}, {Mayer},
  {Baumschlager}, \& {Madau}}]{decataldo24}
{Decataldo}, D., {Shen}, S., {Mayer}, L., {Baumschlager}, B., \& {Madau}, P.
  2024, \aap, 685, A8, \dodoi{10.1051/0004-6361/202346972}

\bibitem[{{Faucher-Gigu{\`e}re} \& {Oh}(2023)}]{faucher-giguere23}
{Faucher-Gigu{\`e}re}, C.-A., \& {Oh}, S.~P. 2023, \araa, 61, 131,
  \dodoi{10.1146/annurev-astro-052920-125203}

\bibitem[{{Field}(1965)}]{field65}
{Field}, G.~B. 1965, \apj, 142, 531, \dodoi{10.1086/148317}

\bibitem[{{Fielding} {et~al.}(2020){Fielding}, {Ostriker}, {Bryan}, \&
  {Jermyn}}]{fielding20}
{Fielding}, D.~B., {Ostriker}, E.~C., {Bryan}, G.~L., \& {Jermyn}, A.~S. 2020,
  \apjl, 894, L24, \dodoi{10.3847/2041-8213/ab8d2c}

\bibitem[{{Fox} {et~al.}(2023){Fox}, {Cashman}, {Kriss}, {de Rosa}, {Plesha},
  {Homayouni}, \& {Richter}}]{fox23}
{Fox}, A.~J., {Cashman}, F.~H., {Kriss}, G.~A., {et~al.} 2023, \apjl, 946, L48,
  \dodoi{10.3847/2041-8213/acc640}

\bibitem[{{Fox} {et~al.}(2014){Fox}, {Wakker}, {Barger}, {Hernandez},
  {Richter}, {Lehner}, {Bland-Hawthorn}, {Charlton}, {Westmeier}, {Thom},
  {Tumlinson}, {Misawa}, {Howk}, {Haffner}, {Ely}, {Rodriguez-Hidalgo}, \&
  {Kumari}}]{fox14}
{Fox}, A.~J., {Wakker}, B.~P., {Barger}, K.~A., {et~al.} 2014, \apj, 787, 147,
  \dodoi{10.1088/0004-637X/787/2/147}

\bibitem[{{Fox} {et~al.}(2016){Fox}, {Lehner}, {Lockman}, {Wakker}, {Hill},
  {Heitsch}, {Stark}, {Barger}, {Sembach}, \& {Rahman}}]{fox16}
{Fox}, A.~J., {Lehner}, N., {Lockman}, F.~J., {et~al.} 2016, \apjl, 816, L11,
  \dodoi{10.3847/2041-8205/816/1/L11}

\bibitem[{{Fox} {et~al.}(2018){Fox}, {Barger}, {Wakker}, {Richter},
  {Antwi-Danso}, {Casetti-Dinescu}, {Howk}, {Lehner}, {D'Onghia}, {Crowther},
  \& {Lockman}}]{fox18}
{Fox}, A.~J., {Barger}, K.~A., {Wakker}, B.~P., {et~al.} 2018, \apj, 854, 142,
  \dodoi{10.3847/1538-4357/aaa9bb}

\bibitem[{{Genel} {et~al.}(2013){Genel}, {Vogelsberger}, {Nelson}, {Sijacki},
  {Springel}, \& {Hernquist}}]{genel13}
{Genel}, S., {Vogelsberger}, M., {Nelson}, D., {et~al.} 2013, \mnras, 435,
  1426, \dodoi{10.1093/mnras/stt1383}

\bibitem[{{Gritton} {et~al.}(2014){Gritton}, {Shelton}, \& {Kwak}}]{gritton14}
{Gritton}, J.~A., {Shelton}, R.~L., \& {Kwak}, K. 2014, \apj, 795, 99,
  \dodoi{10.1088/0004-637X/795/1/99}

\bibitem[{{Gronke} \& {Oh}(2018)}]{gronke18}
{Gronke}, M., \& {Oh}, S.~P. 2018, \mnras, 480, L111,
  \dodoi{10.1093/mnrasl/sly131}

\bibitem[{{Gronke} {et~al.}(2022){Gronke}, {Oh}, {Ji}, \& {Norman}}]{gronke22}
{Gronke}, M., {Oh}, S.~P., {Ji}, S., \& {Norman}, C. 2022, \mnras, 511, 859,
  \dodoi{10.1093/mnras/stab3351}

\bibitem[{{Habe} \& {Ikeuchi}(1980)}]{habe80}
{Habe}, A., \& {Ikeuchi}, S. 1980, Progress of Theoretical Physics, 64, 1995,
  \dodoi{10.1143/PTP.64.1995}

\bibitem[{{Hafen} {et~al.}(2019){Hafen}, {Faucher-Gigu{\`e}re},
  {Angl{\'e}s-Alc{\'a}zar}, {Stern}, {Kere{\v{s}}}, {Hummels}, {Esmerian},
  {Garrison-Kimmel}, {El-Badry}, {Wetzel}, {Chan}, {Hopkins}, \&
  {Murray}}]{hafen19}
{Hafen}, Z., {Faucher-Gigu{\`e}re}, C.-A., {Angl{\'e}s-Alc{\'a}zar}, D.,
  {et~al.} 2019, \mnras, 488, 1248, \dodoi{10.1093/mnras/stz1773}

\bibitem[{{Hayakawa} \& {Fukui}(2024)}]{hayakawa24}
{Hayakawa}, T., \& {Fukui}, Y. 2024, \mnras, 529, 1,
  \dodoi{10.1093/mnras/stae302}

\bibitem[{{Hayward} {et~al.}(2014){Hayward}, {Torrey}, {Springel}, {Hernquist},
  \& {Vogelsberger}}]{hayward14}
{Hayward}, C.~C., {Torrey}, P., {Springel}, V., {Hernquist}, L., \&
  {Vogelsberger}, M. 2014, \mnras, 442, 1992, \dodoi{10.1093/mnras/stu957}

\bibitem[{{Heitsch} {et~al.}(2022){Heitsch}, {Marchal},
  {Miville-Desch{\^e}nes}, {Shull}, \& {Fox}}]{heitsch22}
{Heitsch}, F., {Marchal}, A., {Miville-Desch{\^e}nes}, M.~A., {Shull}, J.~M.,
  \& {Fox}, A.~J. 2022, \mnras, 509, 4515, \dodoi{10.1093/mnras/stab3266}

\bibitem[{{Hobbs} {et~al.}(2013){Hobbs}, {Read}, {Power}, \& {Cole}}]{hobbs13}
{Hobbs}, A., {Read}, J., {Power}, C., \& {Cole}, D. 2013, \mnras, 434, 1849,
  \dodoi{10.1093/mnras/stt977}

\bibitem[{{Hopkins} {et~al.}(2014){Hopkins}, {Kere{\v{s}}}, {O{\~n}orbe},
  {Faucher-Gigu{\`e}re}, {Quataert}, {Murray}, \& {Bullock}}]{fire1}
{Hopkins}, P.~F., {Kere{\v{s}}}, D., {O{\~n}orbe}, J., {et~al.} 2014, \mnras,
  445, 581, \dodoi{10.1093/mnras/stu1738}

\bibitem[{{Hopkins} {et~al.}(2018){Hopkins}, {Wetzel}, {Kere{\v{s}}},
  {Faucher-Gigu{\`e}re}, {Quataert}, {Boylan-Kolchin}, {Murray}, {Hayward},
  {Garrison-Kimmel}, {Hummels}, {Feldmann}, {Torrey}, {Ma},
  {Angl{\'e}s-Alc{\'a}zar}, {Su}, {Orr}, {Schmitz}, {Escala}, {Sanderson},
  {Grudi{\'c}}, {Hafen}, {Kim}, {Fitts}, {Bullock}, {Wheeler}, {Chan},
  {Elbert}, \& {Narayanan}}]{fire2}
{Hopkins}, P.~F., {Wetzel}, A., {Kere{\v{s}}}, D., {et~al.} 2018, \mnras, 480,
  800, \dodoi{10.1093/mnras/sty1690}

\bibitem[{{Hummels} {et~al.}(2019){Hummels}, {Smith}, {Hopkins}, {O'Shea},
  {Silvia}, {Werk}, {Lehner}, {Wise}, {Collins}, \& {Butsky}}]{hummels19}
{Hummels}, C.~B., {Smith}, B.~D., {Hopkins}, P.~F., {et~al.} 2019, \apj, 882,
  156, \dodoi{10.3847/1538-4357/ab378f}

\bibitem[{{Jeffreson} {et~al.}(2021){Jeffreson}, {Keller}, {Winter},
  {Chevance}, {Kruijssen}, {Krumholz}, \& {Fujimoto}}]{jeffreson21}
{Jeffreson}, S. M.~R., {Keller}, B.~W., {Winter}, A.~J., {et~al.} 2021, \mnras,
  505, 1678, \dodoi{10.1093/mnras/stab1293}

\bibitem[{{Kalberla} \& {Haud}(2006)}]{kalberla06}
{Kalberla}, P.~M.~W., \& {Haud}, U. 2006, \aap, 455, 481,
  \dodoi{10.1051/0004-6361:20054750}

\bibitem[{{Kere{\v{s}}} \& {Hernquist}(2009)}]{keres09}
{Kere{\v{s}}}, D., \& {Hernquist}, L. 2009, \apjl, 700, L1,
  \dodoi{10.1088/0004-637X/700/1/L1}

\bibitem[{{Kwak} {et~al.}(2011){Kwak}, {Henley}, \& {Shelton}}]{kwak11}
{Kwak}, K., {Henley}, D.~B., \& {Shelton}, R.~L. 2011, \apj, 739, 30,
  \dodoi{10.1088/0004-637X/739/1/30}

\bibitem[{{Lehner} \& {Howk}(2011)}]{lehner11}
{Lehner}, N., \& {Howk}, J.~C. 2011, Science, 334, 955,
  \dodoi{10.1126/science.1209069}

\bibitem[{{Lehner} {et~al.}(2022){Lehner}, {Howk}, {Marasco}, \&
  {Fraternali}}]{lehner22}
{Lehner}, N., {Howk}, J.~C., {Marasco}, A., \& {Fraternali}, F. 2022, \mnras,
  513, 3228, \dodoi{10.1093/mnras/stac987}

\bibitem[{{Lochhaas} {et~al.}(2023){Lochhaas}, {Tumlinson}, {Peeples},
  {O'Shea}, {Werk}, {Simons}, {Juno}, {Kopenhafer}, {Augustin}, {Wright},
  {Acharyya}, \& {Smith}}]{lochhaas23}
{Lochhaas}, C., {Tumlinson}, J., {Peeples}, M.~S., {et~al.} 2023, \apj, 948,
  43, \dodoi{10.3847/1538-4357/acbb06}

\bibitem[{{Maller} \& {Bullock}(2004)}]{maller04}
{Maller}, A.~H., \& {Bullock}, J.~S. 2004, \mnras, 355, 694,
  \dodoi{10.1111/j.1365-2966.2004.08349.x}

\bibitem[{{McClure-Griffiths} {et~al.}(2009){McClure-Griffiths}, {Pisano},
  {Calabretta}, {Ford}, {Lockman}, {Staveley-Smith}, {Kalberla}, {Bailin},
  {Dedes}, {Janowiecki}, {Gibson}, {Murphy}, {Nakanishi}, \&
  {Newton-McGee}}]{mcclure-griffiths09}
{McClure-Griffiths}, N.~M., {Pisano}, D.~J., {Calabretta}, M.~R., {et~al.}
  2009, \apjs, 181, 398, \dodoi{10.1088/0067-0049/181/2/398}

\bibitem[{{McCourt} {et~al.}(2012){McCourt}, {Sharma}, {Quataert}, \&
  {Parrish}}]{mccourt12}
{McCourt}, M., {Sharma}, P., {Quataert}, E., \& {Parrish}, I.~J. 2012, \mnras,
  419, 3319, \dodoi{10.1111/j.1365-2966.2011.19972.x}

\bibitem[{{Moss} {et~al.}(2013){Moss}, {McClure-Griffiths}, {Murphy}, {Pisano},
  {Kummerfeld}, \& {Curran}}]{moss13}
{Moss}, V.~A., {McClure-Griffiths}, N.~M., {Murphy}, T., {et~al.} 2013, \apjs,
  209, 12, \dodoi{10.1088/0067-0049/209/1/12}

\bibitem[{{Muller} {et~al.}(1963){Muller}, {Oort}, \& {Raimond}}]{muller67}
{Muller}, C.~A., {Oort}, J.~H., \& {Raimond}, E. 1963, Academie des Sciences
  Paris Comptes Rendus, 257, 1661

\bibitem[{{Nelson} {et~al.}(2016){Nelson}, {Genel}, {Pillepich},
  {Vogelsberger}, {Springel}, \& {Hernquist}}]{nelson16}
{Nelson}, D., {Genel}, S., {Pillepich}, A., {et~al.} 2016, \mnras, 460, 2881,
  \dodoi{10.1093/mnras/stw1191}

\bibitem[{{Nelson} {et~al.}(2013){Nelson}, {Vogelsberger}, {Genel}, {Sijacki},
  {Kere{\v{s}}}, {Springel}, \& {Hernquist}}]{nelson13}
{Nelson}, D., {Vogelsberger}, M., {Genel}, S., {et~al.} 2013, \mnras, 429,
  3353, \dodoi{10.1093/mnras/sts595}

\bibitem[{{Nelson} {et~al.}(2019{\natexlab{a}}){Nelson}, {Pillepich},
  {Springel}, {Pakmor}, {Weinberger}, {Genel}, {Torrey}, {Vogelsberger},
  {Marinacci}, \& {Hernquist}}]{nelson19}
{Nelson}, D., {Pillepich}, A., {Springel}, V., {et~al.} 2019{\natexlab{a}},
  \mnras, 490, 3234, \dodoi{10.1093/mnras/stz2306}

\bibitem[{{Nelson} {et~al.}(2019{\natexlab{b}}){Nelson}, {Springel},
  {Pillepich}, {Rodriguez-Gomez}, {Torrey}, {Genel}, {Vogelsberger}, {Pakmor},
  {Marinacci}, {Weinberger}, {Kelley}, {Lovell}, {Diemer}, \&
  {Hernquist}}]{nelson19b}
{Nelson}, D., {Springel}, V., {Pillepich}, A., {et~al.} 2019{\natexlab{b}},
  Computational Astrophysics and Cosmology, 6, 2,
  \dodoi{10.1186/s40668-019-0028-x}

\bibitem[{{Nelson} {et~al.}(2020){Nelson}, {Sharma}, {Pillepich}, {Springel},
  {Pakmor}, {Weinberger}, {Vogelsberger}, {Marinacci}, \&
  {Hernquist}}]{nelson20}
{Nelson}, D., {Sharma}, P., {Pillepich}, A., {et~al.} 2020, \mnras, 498, 2391,
  \dodoi{10.1093/mnras/staa2419}

\bibitem[{{Olano}(2008)}]{olano08}
{Olano}, C.~A. 2008, \aap, 485, 457, \dodoi{10.1051/0004-6361:20077556}

\bibitem[{{Peeples} {et~al.}(2019{\natexlab{a}}){Peeples}, {Behroozi},
  {Bordoloi}, {Brooks}, {Bullock}, {Burchett}, {Chen}, {Chisholm},
  {Christensen}, {Coil}, {Corlies}, {Diamond-Stanic}, {Donahue},
  {Faucher-Gigu{\`e}re}, {Ferguson}, {Fielding}, {Fox}, {French}, {Furlanetto},
  {Gennaro}, {Gilbert}, {Hamden}, {Hathi}, {Hayes}, {Henry}, {Howk}, {Hummels},
  {Kere{\v{s}}}, {Kirby}, {Koekemoer}, {Lan}, {Lanz}, {Law}, {Lehner}, {Lotz},
  {McQuinn}, {McQuinn}, {Munshi}, {Oh}, {O'Meara}, {O'Shea}, {Pacifici},
  {Peek}, {Postman}, {Prescott}, {Putman}, {Quataert}, {Rafelski}, {Ribaudo},
  {Rowlands}, {Rubin}, {Salmon}, {Scarlata}, {Shapley}, {Simons}, {Snyder},
  {Stern}, {Strom}, {Tollerud}, {Tremblay}, {Tripp}, {Tumlinson}, {Tuttle},
  {Bosch}, {Voit}, {Wang}, {Werk}, {Williams}, {Zaritsky}, \&
  {Zheng}}]{peeples19}
{Peeples}, M., {Behroozi}, P., {Bordoloi}, R., {et~al.} 2019{\natexlab{a}},
  \baas, 51, 368, \dodoi{10.48550/arXiv.1903.05644}

\bibitem[{{Peeples} {et~al.}(2019{\natexlab{b}}){Peeples}, {Corlies},
  {Tumlinson}, {O'Shea}, {Lehner}, {O'Meara}, {Howk}, {Earl}, {Smith}, {Wise},
  \& {Hummels}}]{foggie}
{Peeples}, M.~S., {Corlies}, L., {Tumlinson}, J., {et~al.} 2019{\natexlab{b}},
  \apj, 873, 129, \dodoi{10.3847/1538-4357/ab0654}

\bibitem[{{Pillepich} {et~al.}(2018){Pillepich}, {Springel}, {Nelson}, {Genel},
  {Naiman}, {Pakmor}, {Hernquist}, {Torrey}, {Vogelsberger}, {Weinberger}, \&
  {Marinacci}}]{pillepich18}
{Pillepich}, A., {Springel}, V., {Nelson}, D., {et~al.} 2018, \mnras, 473,
  4077, \dodoi{10.1093/mnras/stx2656}

\bibitem[{{Pillepich} {et~al.}(2019){Pillepich}, {Nelson}, {Springel},
  {Pakmor}, {Torrey}, {Weinberger}, {Vogelsberger}, {Marinacci}, {Genel}, {van
  der Wel}, \& {Hernquist}}]{pillepich19}
{Pillepich}, A., {Nelson}, D., {Springel}, V., {et~al.} 2019, \mnras, 490,
  3196, \dodoi{10.1093/mnras/stz2338}

\bibitem[{{Pillepich} {et~al.}(2023){Pillepich}, {Sotillo-Ramos}, {Ramesh},
  {Nelson}, {Engler}, {Rodriguez-Gomez}, {Fournier}, {Donnari}, {Springel}, \&
  {Hernquist}}]{pillepich23}
{Pillepich}, A., {Sotillo-Ramos}, D., {Ramesh}, R., {et~al.} 2023, arXiv
  e-prints, arXiv:2303.16217, \dodoi{10.48550/arXiv.2303.16217}

\bibitem[{{Putman} {et~al.}(2003){Putman}, {Bland-Hawthorn}, {Veilleux},
  {Gibson}, {Freeman}, \& {Maloney}}]{putman03}
{Putman}, M.~E., {Bland-Hawthorn}, J., {Veilleux}, S., {et~al.} 2003, \apj,
  597, 948, \dodoi{10.1086/378555}

\bibitem[{{Putman} {et~al.}(2012){Putman}, {Peek}, \& {Joung}}]{putman12}
{Putman}, M.~E., {Peek}, J.~E.~G., \& {Joung}, M.~R. 2012, \araa, 50, 491,
  \dodoi{10.1146/annurev-astro-081811-125612}

\bibitem[{{Ramesh} \& {Nelson}(2024{\natexlab{a}})}]{ramesh24a}
{Ramesh}, R., \& {Nelson}, D. 2024{\natexlab{a}}, \mnras, 528, 3320,
  \dodoi{10.1093/mnras/stae237}

\bibitem[{{Ramesh} \& {Nelson}(2024{\natexlab{b}})}]{gible}
---. 2024{\natexlab{b}}, \mnras, 528, 3320, \dodoi{10.1093/mnras/stae237}

\bibitem[{{Ramesh} {et~al.}(2024){Ramesh}, {Nelson}, {Fielding}, \&
  {Br{\"u}ggen}}]{ramesh24b}
{Ramesh}, R., {Nelson}, D., {Fielding}, D., \& {Br{\"u}ggen}, M. 2024, arXiv
  e-prints, arXiv:2407.00172, \dodoi{10.48550/arXiv.2407.00172}

\bibitem[{{Ramesh} {et~al.}(2023){Ramesh}, {Nelson}, \& {Pillepich}}]{ramesh23}
{Ramesh}, R., {Nelson}, D., \& {Pillepich}, A. 2023, \mnras, 522, 1535,
  \dodoi{10.1093/mnras/stad951}

\bibitem[{{Richter} {et~al.}(2018){Richter}, {Fox}, {Wakker}, {Howk}, {Lehner},
  {Barger}, {D'Onghia}, \& {Lockman}}]{richter18}
{Richter}, P., {Fox}, A.~J., {Wakker}, B.~P., {et~al.} 2018, \apj, 865, 145,
  \dodoi{10.3847/1538-4357/aadd0f}

\bibitem[{{Richter} {et~al.}(2017){Richter}, {Nuza}, {Fox}, {Wakker}, {Lehner},
  {Ben Bekhti}, {Fechner}, {Wendt}, {Howk}, {Muzahid}, {Ganguly}, \&
  {Charlton}}]{richter17}
{Richter}, P., {Nuza}, S.~E., {Fox}, A.~J., {et~al.} 2017, \aap, 607, A48,
  \dodoi{10.1051/0004-6361/201630081}

\bibitem[{{Sharma} {et~al.}(2012){Sharma}, {McCourt}, {Quataert}, \&
  {Parrish}}]{sharma12}
{Sharma}, P., {McCourt}, M., {Quataert}, E., \& {Parrish}, I.~J. 2012, \mnras,
  420, 3174, \dodoi{10.1111/j.1365-2966.2011.20246.x}

\bibitem[{{Sijacki} {et~al.}(2012){Sijacki}, {Vogelsberger}, {Kere{\v{s}}},
  {Springel}, \& {Hernquist}}]{sijacki12}
{Sijacki}, D., {Vogelsberger}, M., {Kere{\v{s}}}, D., {Springel}, V., \&
  {Hernquist}, L. 2012, \mnras, 424, 2999,
  \dodoi{10.1111/j.1365-2966.2012.21466.x}

\bibitem[{{Smoker} {et~al.}(2011){Smoker}, {Fox}, \& {Keenan}}]{smoker11}
{Smoker}, J.~V., {Fox}, A.~J., \& {Keenan}, F.~P. 2011, \mnras, 415, 1105,
  \dodoi{10.1111/j.1365-2966.2011.18647.x}

\bibitem[{{Spitzer}(1956)}]{spitzer56}
{Spitzer}, Lyman, J. 1956, \apj, 124, 20, \dodoi{10.1086/146200}

\bibitem[{{Springel}(2010)}]{springel10}
{Springel}, V. 2010, \mnras, 401, 791, \dodoi{10.1111/j.1365-2966.2009.15715.x}

\bibitem[{{Suresh} {et~al.}(2015){Suresh}, {Bird}, {Vogelsberger}, {Genel},
  {Torrey}, {Sijacki}, {Springel}, \& {Hernquist}}]{suresh15}
{Suresh}, J., {Bird}, S., {Vogelsberger}, M., {et~al.} 2015, \mnras, 448, 895,
  \dodoi{10.1093/mnras/stu2762}

\bibitem[{{Suresh} {et~al.}(2019){Suresh}, {Nelson}, {Genel}, {Rubin}, \&
  {Hernquist}}]{suresh19}
{Suresh}, J., {Nelson}, D., {Genel}, S., {Rubin}, K. H.~R., \& {Hernquist}, L.
  2019, \mnras, 483, 4040, \dodoi{10.1093/mnras/sty3402}

\bibitem[{{Tan} {et~al.}(2021){Tan}, {Oh}, \& {Gronke}}]{tan21}
{Tan}, B., {Oh}, S.~P., \& {Gronke}, M. 2021, \mnras, 502, 3179,
  \dodoi{10.1093/mnras/stab053}

\bibitem[{{Tan} {et~al.}(2023){Tan}, {Oh}, \& {Gronke}}]{tan23}
---. 2023, \mnras, 520, 2571, \dodoi{10.1093/mnras/stad236}

\bibitem[{{Tripp}(2022)}]{tripp22}
{Tripp}, T.~M. 2022, \mnras, 511, 1714, \dodoi{10.1093/mnras/stac044}

\bibitem[{{Tumlinson} {et~al.}(2017){Tumlinson}, {Peeples}, \&
  {Werk}}]{tumlinson17}
{Tumlinson}, J., {Peeples}, M.~S., \& {Werk}, J.~K. 2017, \araa, 55, 389,
  \dodoi{10.1146/annurev-astro-091916-055240}

\bibitem[{{van de Voort} {et~al.}(2019){van de Voort}, {Springel}, {Mandelker},
  {van den Bosch}, \& {Pakmor}}]{vandevoort19}
{van de Voort}, F., {Springel}, V., {Mandelker}, N., {van den Bosch}, F.~C., \&
  {Pakmor}, R. 2019, \mnras, 482, L85, \dodoi{10.1093/mnrasl/sly190}

\bibitem[{{van Woerden} {et~al.}(1957){van Woerden}, {Rougoor}, \&
  {Oort}}]{vanwoerden57}
{van Woerden}, H., {Rougoor}, G.~W., \& {Oort}, J.~H. 1957, Academie des
  Sciences Paris Comptes Rendus, 244, 1691

\bibitem[{{Vogelsberger} {et~al.}(2012){Vogelsberger}, {Sijacki},
  {Kere{\v{s}}}, {Springel}, \& {Hernquist}}]{vogelsberger12}
{Vogelsberger}, M., {Sijacki}, D., {Kere{\v{s}}}, D., {Springel}, V., \&
  {Hernquist}, L. 2012, \mnras, 425, 3024,
  \dodoi{10.1111/j.1365-2966.2012.21590.x}

\bibitem[{{Voit}(2021)}]{voit21}
{Voit}, G.~M. 2021, \apjl, 908, L16, \dodoi{10.3847/2041-8213/abe11f}

\bibitem[{{Wakker} \& {van Woerden}(1997)}]{wakker97}
{Wakker}, B.~P., \& {van Woerden}, H. 1997, \araa, 35, 217,
  \dodoi{10.1146/annurev.astro.35.1.217}

\bibitem[{{Weinberger} {et~al.}(2017){Weinberger}, {Springel}, {Hernquist},
  {Pillepich}, {Marinacci}, {Pakmor}, {Nelson}, {Genel}, {Vogelsberger},
  {Naiman}, \& {Torrey}}]{weinberger17}
{Weinberger}, R., {Springel}, V., {Hernquist}, L., {et~al.} 2017, \mnras, 465,
  3291, \dodoi{10.1093/mnras/stw2944}

\bibitem[{{Westmeier}(2018)}]{westmeier18}
{Westmeier}, T. 2018, \mnras, 474, 289, \dodoi{10.1093/mnras/stx2757}

\end{thebibliography}
\bibliographystyle{aasjournal}


\end{document}